\documentclass{article}

\PassOptionsToPackage{numbers, compress}{natbib}

\usepackage[preprint]{neurips_2026}

\usepackage[utf8]{inputenc} 
\usepackage[T1]{fontenc}    
\usepackage{hyperref}       
\hypersetup{hypertexnames=false}
\usepackage{url}            
\usepackage{booktabs}       
\usepackage{amsfonts}       
\usepackage{nicefrac}       
\usepackage{microtype}      
\usepackage{xcolor}         
\usepackage[numbers]{natbib}
\usepackage{colortbl}
\usepackage{amsmath,amssymb,amsfonts}
\usepackage{amsthm}
\usepackage{mathtools}
\usepackage{bm}
\usepackage{enumitem}
\usepackage{algorithm}
\usepackage{algpseudocode}
\usepackage{multirow, multicol}
\usepackage{longtable}
\usepackage{listings}
\usepackage{upquote}
\usepackage{fontawesome5}
\usepackage{wrapfig}
\usepackage{adjustbox}

\newtheorem{definition}{Definition}
\newtheorem{remark}{Remark}
\DeclareMathOperator*{\argmax}{arg\,max}

\definecolor{candoBlue}{RGB}{70,130,180}
\definecolor{ricoColor}{HTML}{E6F4EA}
\definecolor{publayColor}{HTML}{F5F0E6}
\lstdefinestyle{candoprompt}{
	basicstyle=\ttfamily\scriptsize,
	breaklines=true,
	breakatwhitespace=false,
	columns=fullflexible,
	keepspaces=true,
	showstringspaces=false,
	upquote=true,
	literate=
		{°}{{$^\circ$}}1
		{—}{{---}}3
		{“}{{``}}2
		{”}{{''}}2
		{→}{{$\rightarrow$}}1,
	frame=single,
	xleftmargin=0.5em,
	xrightmargin=0.5em,
	aboveskip=0.5\baselineskip,
	belowskip=0.5\baselineskip
}

\title{%
  \begin{tabular}{@{}c@{\hspace{0.2em}}c@{}}
    \adjustbox{valign=c}{\includegraphics[height=2.5em]{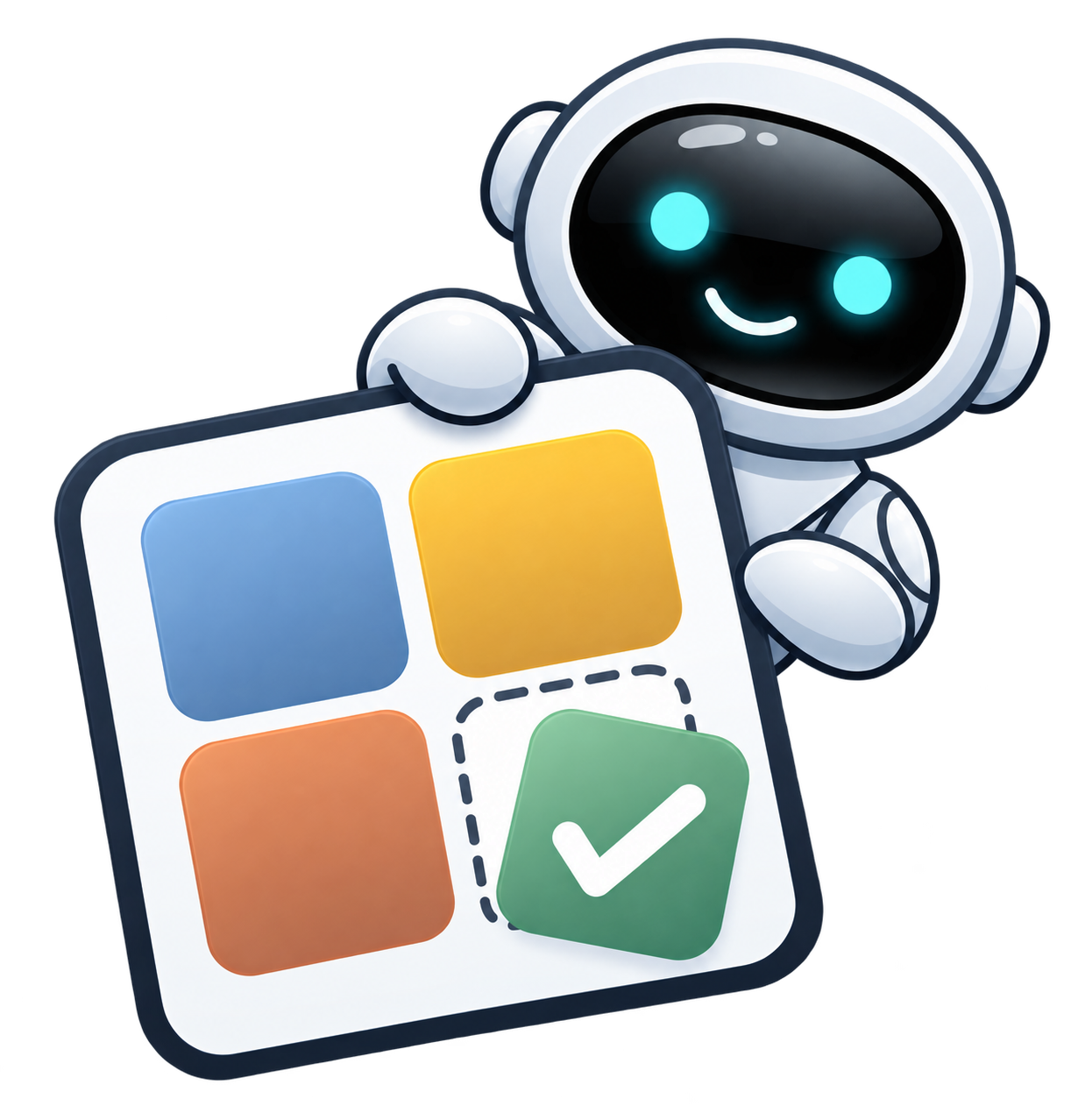}} &
    \begin{tabular}{@{}c@{}}
      CANDO: \underline{C}ooperative \underline{A}gentic \underline{N}etwork for \\
      Layout \underline{D}esign \underline{O}ptimization
    \end{tabular}
  \end{tabular}%
}

\author{
    Athanasios Masouris$^{1}$
    \quad
    Jing Zheng$^{2}$
    \quad
    Benjamin Chandler$^{1}$
    \quad
    Hadi Jamali\mbox{-}Rad$^{1,3}$ \\[0.5em]
    $^{1}$ Shell Information Technology International,
    $^{2}$ Shell China Limited\\
    $^{3}$ Delft University of Technology (TU Delft)
}

\begin{document}

\maketitle

\begin{abstract}\vspace{-0.2cm}
Layout generation for real-world facilities is a challenging problem, requiring reasoning over irregular site boundaries, heterogeneous orientations, access-aware placements, and motion-planning feasibility. Yet, most existing layout benchmarks in the generative AI space target simpler placements over rectangular domains and rely on distributional metrics such as FID and IoU that reward conformity to dataset priors, thus discounting design innovation. Motivated by these gaps, we introduce \textbf{ALPS-Bench}, a benchmark of $1,000$ professionally annotated real-world facility layouts paired with an instance-specific scoring protocol grounded in a structured design manual. As a strong baseline for \textbf{ALPS-Bench}, we propose \texttt{CANDO}, a training-free multi-agent framework in which specialized agents iteratively refine layouts through a verification-grounded loop, concentrating reasoning on strategic spatial decisions. We demonstrate that \texttt{CANDO} surpasses state-of-the-art trained and LLM-based baselines on the widely adopted \textbf{PubLayNet}, \textbf{RICO}, and \textbf{PKU-PosterLayout} benchmarks, establishing cooperative agentic design as a broadly effective recipe for constraint-aware layout synthesis.
\end{abstract}

\begin{center}
\small
\href{https://shell-oss.github.io/cando/}{\faGlobe\ \textbf{Project Page}}
\qquad
\href{https://github.com/Shell-OSS/cando}{\faGithub\ \textbf{Codebase}}
\qquad
\href{https://github.com/Shell-OSS/cando/tree/main/data}{\faDatabase\ \textbf{ALPS-Bench}}
\end{center}

\section{Introduction}\label{sec:intro}\vspace{-0.2cm}
Layout generation has recently received an upsurge of attention as a subfield of generative modeling, with strong results reported across floor planning \citep{wu2019data, kalervo2019cubicasa5k,van2024msd,abouagour2025resplan}, document design \citep{zhong2019publaynet, pfitzmann2022doclaynet}, poster design \citep{hsu2023posterlayout}, and graphical user interfaces synthesis \citep{deka2017rico, wu2023webui} tasks. Yet the regime captured by these interesting benchmarks is notably simpler than real-world design practice: floor-planning datasets are dominated by indoor, largely axis-aligned room arrangements; document and web layouts reduce to rectangular composition over discrete canvases. Besides, their evaluation strategy is predominantly distributional, with metrics such as FID, IoU, SSIM, and graph similarity rewarding resemblance to the dataset prior. The consequence is twofold. First, design innovation is implicitly penalized; i.e., valid yet atypical configurations are scored down simply for departing from the empirical distribution, collapsing the task into prior matching. Second, methods that excel under these protocols rarely transfer to settings where compositionality, orientation, and inter-element spatial relations are first-class concerns rather than emergent regularities of the dataset.

A canonical real-world instance of this overlooked regime is asset layout design: arranging service units, parking bays, access points, and auxiliary buildings within an operational facility. Unlike its document and floor-plan counterparts, this problem requires reasoning over irregular, instance-specific site boundaries; enforcing orientation consistency across heterogeneous functional groups; organizing elements with respect to designated access points; and certifying motion-planning feasibility for multiple vehicle scenarios traversing the site. These requirements are entangled, namely: a placement that satisfies non-overlap may still block a routing scenario, and a compositionally clean clustering may violate access-aware constraints. We find empirically that neither distributional metrics nor (well-equipped) single-agent baselines can adequately address this challenging problem. More specifically, monolithic generators struggle on precisely the joint demands of spatial arithmetic, multi-constraint reasoning, and long-horizon planning that the problem entails, pointing to the need for an explicitly decomposed, verification-grounded solution architecture. Intrigued by these gaps and challenges, we make the following three core contributions: 
\begin{itemize}
    \item[I.] We introduce the \underline{A}sset \underline{L}ayout \underline{P}lanning and \underline{S}ynthesis Benchmark (\textbf{ALPS-Bench}): 1,000 real-asset professionally-annotated layouts, each paired with a structured design manual of diverse spatial constraints. Our proposed evaluation protocol scores compliance across a variety of design criteria for standardized evaluation on a held-out test set.
    \item[II.] We propose \texttt{CANDO}, to the best of our knowledge, one of the first cooperative agentic AI frameworks for constraint-aware layout synthesis and a strong baseline for \textbf{ALPS-Bench}. \texttt{CANDO} offers several innovative design choices, including structured cooperation among agents, tool-grounded verification, and motion planning.
    \item[III.] We demonstrate that even a simplified version of \texttt{CANDO} achieves state-of-the-art mIoU and alignment on the widely adopted \textbf{PubLayNet} and \textbf{RICO} benchmarks, and leading overlap and strict-underlay performance on \textbf{PKU-PosterLayout}, corroborating that cooperative agentic frameworks are a broadly effective recipe for layout generation.
\end{itemize}
\section{Related Work}\label{sec:related-work}\vspace{-0.2cm}
\paragraph{Layout Design and Its Benchmarks.} Automated layout generation has been studied across a range of domains, including graphical user interfaces~\citep{deka2017rico}, document design~\citep{zhong2019publaynet}, content-aware poster design~\citep{hsu2023posterlayout}, indoor scene arrangement~\citep{paschalidou2021atiss,tang2024diffuscene}, and architectural floor planning~\citep{nauata2021house,shabani2023housediffusion}. Early work relied on GANs~\citep{li2019layoutgan} and VAEs~\citep{jyothi2019layoutvae, lee2020neural}, achieving reasonable qualitative results but at the cost of task-specific training on large labeled datasets. Transformer-based methods then broadened the scope, with LayoutTransformer~\citep{gupta2021layouttransformer} enabling layout completion via self-attention, and LayoutFormer++~\citep{jiang2023layoutformer++} unifying diverse constraint types into a single sequence format. Diffusion models followed, with LayoutDM~\citep{inoue2023layoutdm}, LayoutFlow~\citep{guerreiro2024layoutflow}, and DiffuScene~\citep{tang2024diffuscene} pushing generation quality further. These methods require training and are evaluated on distributional metrics like FID and overlap that reward proximity to the training prior rather than actual constraint satisfaction. Large language models (LLMs) have offered a different path. LayoutPrompter~\citep{lin2023layoutprompter} and LayoutGPT~\citep{feng2023layoutgpt} pioneered training-free layout generation via in-context learning (ICL), while fine-tuned variants such as LayoutNUWA~\citep{tang2023layoutnuwa} and PosterLlama~\citep{seol2024posterllama} improved quality by representing layouts as structured HTML code. More recently, LayoutCoT~\citep{shi2025layoutcot} incorporated chain-of-thought (CoT) reasoning, while LaRa~\citep{laragenerative} introduced reinforcement learning (RL) from verifiable rewards. An emerging agentic direction pushes this further, with DisCo-Layout~\citep{gao2025disco} and AutoLayout~\citep{chen2025autolayout} proposing multi-agent loops for iterative spatial refinement. Notably, RICO~\citep{deka2017rico} and PubLayNet~\citep{zhong2019publaynet} remain the predominant benchmarks adopted by almost all of the above, focusing on much simpler tasks and relying on prior matching metrics (such as mIoU and FID), leaving a gap in the literature that jointly motivated the proposed \textbf{ALPS-Bench} and \texttt{CANDO}.

\paragraph{Cooperative Agentic AI.} The structured reasoning capabilities of LLMs through CoT have been further enhanced by methods that guide such models to verify and refine their own responses, including Self-Refine~\citep{madaan2023self} and Reflexion~\citep{shinn2023reflexion}. ReAct~\citep{yao2022react} established the interplay between structured reasoning and environment-grounded actions, paving the way for the modern agentic systems. Toolformer~\citep{schick2023toolformer} demonstrated that LLMs benefit from offloading computations to external tools, which is particularly important given their known spatial arithmetic limitations~\citep{chen2024spatialvlm}. Frameworks, such as Generative Agents~\citep{park2023generative}, MemGPT~\citep{packer2024memgptllmsoperatingsystems}, and A-Mem~\citep{xu2025mem}, further showed that episodic memory management is essential for coherent long-horizon behavior. Building upon these findings, multi-agent orchestration frameworks, such as AutoGen~\citep{wu2024autogen}, MetaGPT~\citep{hong2023metagpt}, CAMEL~\citep{li2023camel}, AgentVerse~\citep{chen2023agentverse}, and MINDcraft~\citep{white2025blocks}, establish that role-specialized cooperating agents consistently outperform monolithic models on complex tasks~\citep{du2024improving}. Inspired by this prior art, we propose \texttt{CANDO}, a novel approach to constraint-aware layout design consisting of a cooperative agentic network that leverages tool-grounded verification, episodic feedback memory, role-specialized cooperation, and block-wise BoN \citep{beirami2024theoretical} (a.k.a CoDe \citep{singh2025code}) into a single training-free framework.
\section{Problem Formulation}\label{sec:formulation}\vspace{-0.2cm}

We formalize layout synthesis as constrained optimization of rigid-body placement over a $2$D plane. The goal is to determine the position and  orientation of each element so as to satisfy geometric feasibility requirements and minimize violation of design constraints.

\paragraph{Layout Domain.} It is comprised of the layout premise $\mathcal{P}\subset\mathbb{R}^2$; an empty outer polygon within which the elements are to be placed. And, a set of access points (i.e., \textsc{entrance}, \textsc{exit}) sitting on the edges of the premise $\mathcal{P}$. Access points define designated regions through which vehicles may enter or leave the domain, and they take a pivotal role in operational constraints such as routing feasibility.

\paragraph{Elements Configuration.} Elements $e_i$ ($i \in [E]$) are to be placed within the premise $\mathcal{P}$. Each $e_i = (c_i, G_i)$ has a semantic class $c_i\in\mathcal{C}$ and a fixed canonical shape/body $G_i\subset\mathbb{R}^2$, a compact region centered at the origin (e.g. a rectangle). We denote $\mathcal{E} = \{(c_i,\,G_i)\}_{i=1}^{E}$ for the full element specification. Element shapes/sizes, classes, and counts are \emph{fixed inputs}; only their spatial placement is optimized. See Fig.~\ref{fig:layout-sample} for an example layout illustrating premise, access points and elements.

Each $e_i$ is displaced by assigning a centroid $\mathbf{p}_i=(x_i,y_i)\in\mathbb{R}^2$ and an orientation angle $\theta_i\in[0,2\pi)$. The displaced element is the element canonical shape $G_i$ rotated by~$\theta_i$ and translated to~$\mathbf{p}_i$ following:
\begin{equation}\label{eq:placed}
  \Omega_i(\mathbf{p}_i,\theta_i)
    = R(\theta_i)\,G_i + \mathbf{p}_i,
  \qquad
  R(\theta_i)=
  \begin{pmatrix}
    \cos\theta_i & -\sin\theta_i \\
    \sin\theta_i & \phantom{-}\cos\theta_i
  \end{pmatrix}.
\end{equation}
For the sake of simplicity, we omit the fixed canonical shape $G_i$ and class label $c_i$ in denoting the \emph{displaced elements} $\Omega_i$ resorting only to the parameters $(\mathbf{p}_i,\theta_i)$. We bundle these tuples in the \emph{elements configuration} $\mathbf{q}=\{(\mathbf{p}_i,\theta_i)\}_{i=1}^{E} \in \mathcal{Q}$, with $\mathcal{Q}$ being the set of all possible configurations. As a result, $\mathbf{q}$ now can serve as our optimization parameter, focusing only on the variable parts of the elements and not the layout domain (premise and access points).

\paragraph{Constraints.}
Design manual $\mathcal{M} = \{\mathcal{HC}, \mathcal{SC}\}$ covers two sets of desired specifications:

\emph{Hard constraints} ($\mathcal{HC}$) encode three necessary conditions that yield \emph{valid} configurations $\mathcal{Q}_\textrm{valid} \subset \mathcal{Q}$: \vspace{-0.15cm}
\begin{enumerate}[label=(\roman*),nosep,leftmargin=*]
  \item \textbf{Containment.} Every placed element lies within the premise: $\Omega_i\subseteq\mathcal{P}$, ~$\forall i\in[E]$.
  \item \textbf{Non-overlap.} Distinct elements do not intersect or fully overlap: $\Omega_i \cap\, \Omega_j =\emptyset$,~$\forall i\neq j$.
  \item \textbf{Routing feasibility.} A finite set of predefined scenarios $\mathcal{R}$ specifies required vehicle routes as ordered sequences of intermediate points (e.g., [\textsc{vehicle type}, $\textsc{entrance}\to\textsc{waypoint}\to\cdots\to\textsc{exit}$]). For every scenario and every concrete instance of intermediate points, a collision-free path must exist through the free space $\mathcal{P}\setminus\bigcup_{i}\Omega_i$.
\end{enumerate}

\emph{Soft constraints} ($\mathcal{SC}$) quantify how well a configuration satisfies the following design properties: \vspace{-0.15cm}
\begin{enumerate}[label=(\roman*),nosep,leftmargin=*]
  \item \textbf{Compositionality.} Prescribes arrangement patterns and intra-group distance thresholds.
  \item \textbf{Orientation.} Defines permitted orientations and intra-group element-wise alignment.
  \item \textbf{Proximity.} Encodes stacking relations, clearance zones, and inter-element distance bounds.
\end{enumerate}

\paragraph{Objective.}
The goal is to find a valid configuration that maximizes total weighted score:
\begin{equation}\label{eq:objective}
  \boxed{\;
  \mathbf{q}^{\star}
    =\argmax_{\mathbf{q}\,\in\,\mathcal{Q}_{\mathrm{valid}}}
      \;\mathcal{J}(\mathbf{q})
    \;\triangleq\;
      \sum_{s=1}^{S}w_s\,f_s(\mathbf{q}),
  \;}
\end{equation}
with $S = |\mathcal{SC}|$, and scoring function $f_s \in [0,1]$ indicating compliance against soft constraint $s \in \mathcal{SC}$ when $f_s(\mathbf{q})=1$. Scoring weights sum up to $\sum_{s}w_s=1$. A configuration is considered \emph{compliant} if it is valid (satisfying all hard constraints $\mathcal{HC}$) and $\mathcal{J}(\mathbf{q})=1$.

\begin{definition}[Input requirements]\label{def:instance}
A layout input requirement is a tuple $\mathcal{I} =(\mathcal{P},\;\mathcal{A},\;\mathcal{E},\mathcal{M})$, specifying the layout premise, access points, element set, and constraint partition.
\end{definition}

\begin{remark}[Generalization to higher dimensions]\label{rem:3d}
The formulation extends naturally to volumetric layout synthesis (e.g., multi-floor facility planning) by extending Eq.~\ref{eq:placed} to its 3-D counterpart \citep{gower1975generalized}.
\end{remark}
\section{Asset Layout Planning and Synthesis (ALPS) Benchmark}\label{sec:dataset}\vspace{-0.2cm}

As discussed in Section~\ref{sec:intro}, prior layout benchmarks do not fully capture the geometric and operational intricacies of real-life complex design problems such as outdoor facility/asset design, and are typically evaluated by similarity scores against dataset priors rather than constraint satisfaction.
\begin{figure}[t]
    \centering
    \includegraphics[width=\textwidth, trim=0 0 0 0.5cm, clip]{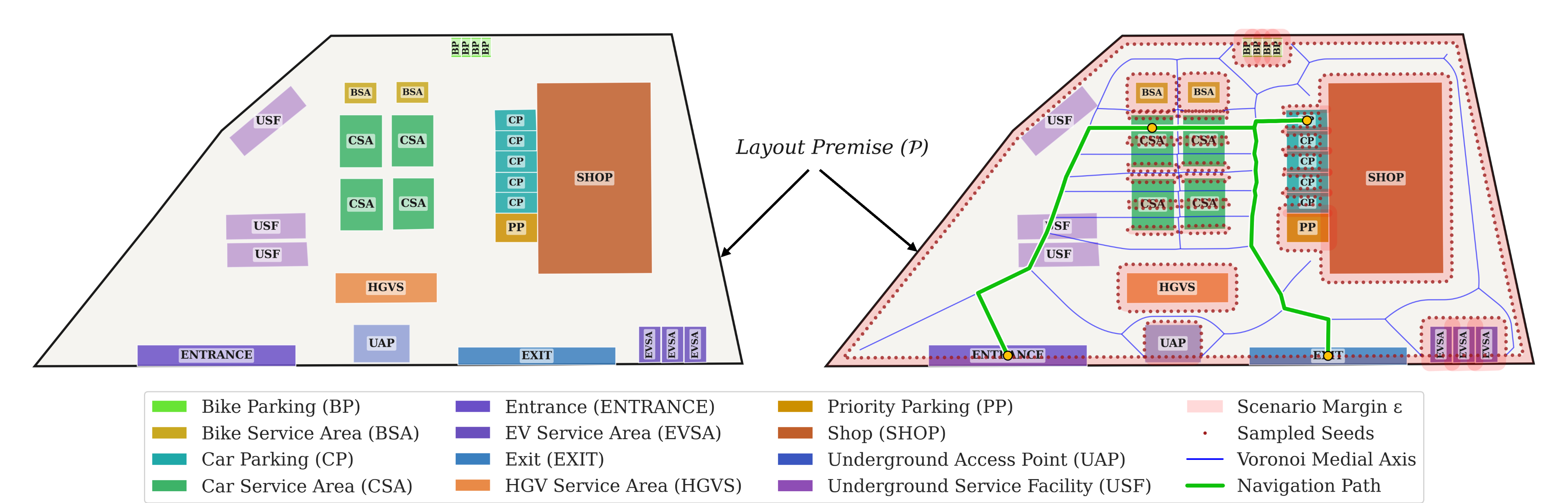}\vspace{-0.3cm}
    \caption{\textbf{Left}: Sample from \textbf{ALPS-Bench}. \textbf{Right}: Voronoi-based motion planning.}
    \label{fig:layout-sample}
\end{figure}
Intrigued by these gaps, we introduce the \underline{A}sset \underline{L}ayout \underline{P}lanning and \underline{S}ynthesis Benchmark (\textbf{ALPS-Bench}), a collection of $1{,}000$ annotated layouts from real operational sites. Polygon-level annotations were produced by professional annotators working directly from original CAD design drawings of each facility, yielding geometrically accurate element boundaries that reflect the as-built state of the sites. Each layout is composed of elements drawn from $14$ semantic classes organized into four functional groups: (i) site premise and access: referring to layout premise $\mathcal{P}$, \textsc{entrance} and \textsc{exit}; (ii) service infrastructure: three service unit variants differentiated by vehicle size (\textsc{csa}, \textsc{bsa}, \textsc{hgvs}), an electric vehicle charging area (\textsc{evsa}), an underground service facility (\textsc{usf}), and an underground access point (\textsc{uap}); (iii) building and auxiliary services: a retail building (\textsc{shop}); and (iv) parking: four bay variants: car (\textsc{cp}), bicycle (\textsc{bp}), priority (\textsc{pp}), and heavy-goods vehicle (\textsc{hgvp}). Element composition is instance-specific: not all $14$ classes appear in every layout, introducing compositional variability and class imbalance characteristic of real facility layout design. An example layout is illustrated in Fig.~\ref{fig:layout-sample} (left). The layouts are governed by a shared design manual $\mathcal{M}$ (as discussed in Section~\ref{sec:formulation}, with details in Appendix~\ref{app:design-manual}), released alongside the dataset. The full routing scenarios and soft constraint specification are detailed in Appendices~\ref{app:routing} and~\ref{app:soft-constraints}. In contrast to existing web and document layout benchmarks, \textbf{ALPS-Bench} offers challenging design through variable premise shape, varying numbers of elements and access points, and an imbalanced class distribution.

\textbf{Data Split and Structure.} The $1{,}000$ layouts are partitioned into a development set of $900$ instances ($\mathcal{D}^{\textup{dev}}$) and a held-out test set of $100$ instances ($\mathcal{D}^{\textup{test}}$). Development instances are released with full annotations (input requirements $\mathcal{I}$ together with the ground-truth configuration $\mathbf{q}^{\dagger}$) making this set suitable for training, in-context learning, and hyperparameter selection. For test instances, only the input requirements $\mathcal{I}$ are made public; ground-truth configurations are withheld to support standardized automated evaluation on a hosted online server. Further details on dataset statistics and annotation format can be found in Appendix~\ref{app:data-statistics} and~\ref{app:annotation-format}, respectively.

\subsection{Evaluation and Benchmarking}
\label{ssec:eval}

To evaluate newly designed layouts, we propose an instance-specific scorer instead of metrics such as IoU or FID, which can penalize design innovation by promoting prior matching. The scoring function evaluates generated instances from two perspectives: (i) satisfaction of design manual ($\mathcal{M} = \{\mathcal{HC}, \mathcal{SC}\}$); and (ii) comparison against withheld ground-truth layouts $\mathbf{q}^\dagger$ in terms of routing efficiency. Let us delve deeper into the evaluation metrics.

\paragraph{Hard Constraints Metrics (HCM).}
To evaluate satisfaction of the hard constraints ({$\mathcal{HC}$}) of the design manual ($\mathcal{M}$) in Section~\ref{sec:formulation}, this metric investigates the existence and correct sizing of all elements specified by input requirement $\mathcal{I}$; containment ($\Omega_i \subseteq \mathcal{P}$, $\forall i$); non-overlap ($\Omega_i \cap \Omega_j = \emptyset$, $\forall i \neq j$); and routing feasibility across scenarios in $\mathcal{R}$. While the other conditions are straightforward to measure, routing feasibility is a non-trivial assessment task. To assess unobstructed navigation for a routing scenario, we adopt a simple Voronoi-based motion-planning approach~\citep{latombe2012robot, choset1995sensor}. We first inflate the borders of obstacles (i.e., $e_i \notin \mathcal{R}$) by a scenario-dependent safety margin $\epsilon$. We then sample seed points on the inflated borders and use them to construct a Voronoi diagram. The Voronoi medial axes are used to identify paths with maximum clearance. This is shown in Fig.~\ref{fig:layout-sample} (right), with further details in Appendix~\ref{app:voronoi}. Missing or incorrectly sized elements result in HCM~$= 0$. Otherwise, the score is a continuous value in $(0, 1]$, computed as $\text{HCM} = \alpha_\text{g} \times \text{HC}_\text{g} + \alpha_\text{f} \times \text{HC}_\text{f}$, where the geometric validity score HC\textsubscript{g} proportionally penalizes out-of-boundary and overlapping areas, and the flow score HC\textsubscript{f} is the fraction of feasible routing scenarios. We use $\alpha_{\text{g}} = 0.2$ and $\alpha_{\text{f}} = 0.8$, assigning more weight to valid routing because it is considerably more complex to address.

\paragraph{Soft Constraint Metrics (SCM).}
To evaluate satisfaction of the soft constraints ({$\mathcal{SC}$}) part of the design manual ($\mathcal{M}$), this metric investigates three aspects. Proximity (PR) $\in [0, 1]$ checks stacking/adjacency, minimum clearance, and maximum inter-element distance requirements, penalizing each violation at $-0.1$. Compositionality (CO) $\in [0, 1]$ checks whether elements of each semantic class form the spatial formations required by the design manual (e.g., line arrays, matrix grids), also penalizing fragmentation and wrong formations at $-0.1$ per violation. Orientation (OR) $\in [0, 1]$ checks intra-group angular alignment, with penalties scaling continuously with angular deviation. The score is a continuous value in $[0,1]$, computed as $\text{SCM} = \alpha_{\text{p}} \times \text{PR} + \alpha_{\text{c}} \times \text{CO} + \alpha_{\text{o}} \times \text{OR}$. We currently use $ \alpha_{\text{p}}=0.6, ~\alpha_{\text{c}}=0.3$ and  $\alpha_{\text{o}}=0.1$, reflecting the proportional number of soft constraint rules for each category included in $\mathcal{M}$.

\paragraph{Ground Truth Metrics (GTM).} To go beyond constraint satisfaction, this metric compares generated layouts against the withheld ground-truth configurations $\mathbf{q}^{\dagger}$ along three operational dimensions: path length (PL), maneuverability (MN), and path clearance (PC), each normalized to $[0,1]$ with $0.5$ indicating equivalence to the ground truth. PL is scored as $\overline{\ell}_{\mathrm{gt}} / (\overline{\ell}_{\mathrm{gt}} + \overline{\ell}_{\mathrm{gen}})$; MN as $\overline{n}_{\mathrm{gt}} / (\overline{n}_{\mathrm{gt}} + \overline{n}_{\mathrm{gen}})$, where $\overline n$ is the average number of turns; and PC as $\overline{c}_{\mathrm{gen}} / (\overline{c}_{\mathrm{gt}} + \overline{c}_{\mathrm{gen}})$, where $\overline{c}$ is the average minimum distance to obstacles along the paths. Subscripts $(.)_\mathrm{gt}$ and $(.)_\mathrm{gen}$ refer to ground-truth and generated layouts, respectively. Values above $0.5$ indicate shorter routes, fewer turns, or wider clearances relative to the ground truth. Invalid scenarios receive $\text{PL} = \text{MN} = \text{PC} =0$.

\paragraph{Aggregate Score (AGG).}
With all sub-metrics bounded in $[0, 1]$, we define the final scores:
\begin{equation}
    \text{AGG} = \tfrac{1}{2}(\text{CS} + \text{RE}), ~
    \text{CS} =
    \begin{cases}
    \text{SCM} & \text{if } \text{HCM} = 1 \\ \alpha \times \text{HCM} \times \text{SCM} & \text{if } \text{HCM} < 1  \end{cases}, ~
    \text{RE} = \tfrac{1}{3}(\text{PL} + \text{MN} + \text{PC}).
    \label{eq:agg_metrics}
\end{equation}
Our constraint satisfaction metrics are summarized in CS, where invalid final layouts ($\text{HCM} < 1$) are penalized with a capped score of $\alpha$. We set $\alpha = 0.5$ to emphasize that invalid layouts are undesirable. The routing efficiency metrics are summarized in RE, and CS and RE are averaged to produce the aggregate score (AGG). All scores are computed per layout and then averaged over the relevant split.
\begin{figure}[t]
\centering
\includegraphics[width=\linewidth]{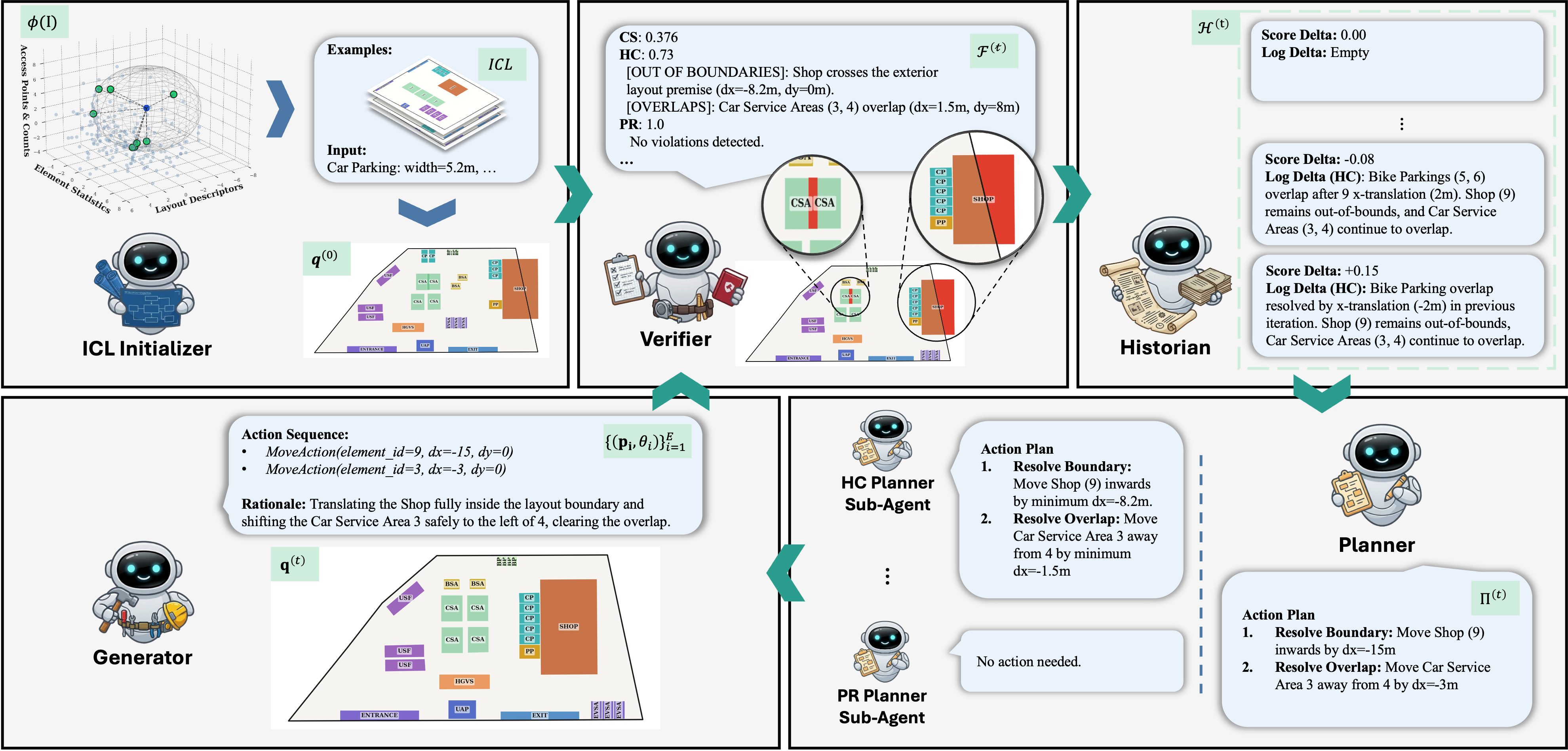}
\vspace{-0.5cm}
\caption{The interplay and cooperation among the agents in \texttt{CANDO}. Upon an ICL Initialization, Verifier, Historian, Planner and Generator communicate in a structured manner within a loop.} \label{fig:methodology}
\end{figure}
\section{Methodology}\label{sec:methodology}\vspace{-0.2cm}
To solve the optimization problem introduced in Section~\ref{sec:formulation} and instantiated with \textbf{ALPS-Bench}, we propose \underline{C}ooperative \underline{A}gentic \underline{N}etwork for Layout \underline{D}esign \underline{O}ptimization (\texttt{CANDO}), a multi-agent framework for constraint-aware layout synthesis. As motivated in Section~\ref{sec:intro}, this cooperative architecture distributes reasoning tasks across specialized agents to avoid the failure modes of single-model generation and to enable processing with resource-constrained agents (with smaller backbones). This design choice is supported by both prior art \citep{li2023camel, hong2023metagpt, wu2024autogen, wang2024mixture} and our experiments in Section~\ref{sec:experiments}. The proposed solution operates in two phases. First, an \emph{in-context learning initializer (ICL Initializer)} produces a seed configuration $\mathbf{q}^{(0)}$ by conditioning an LLM on structurally similar layouts retrieved from the development data. Second, four specialized agents (\emph{Verifier}, \emph{Historian}, \emph{Planner}, and \emph{Generator}) cooperate to refine $\mathbf{q}^{(0)}$ toward $\mathbf{q}^{\star}$ through a parallel-branching iterative loop adapted from CoDe's blockwise sampling scheme for diffusion models~\citep{singh2025code}. Crucially, since language models struggle with spatial understanding and coordinate arithmetic~\citep{yamada2023evaluating}, \texttt{CANDO} deliberately offloads low-level geometric computations to deterministic tools.

The two-phase design reflects the asymmetric interplay between feasibility and design quality, as hard-constraint violations would render soft-constraint quality irrelevant. The proposed solution design and inner-mechanics of the agentic cooperation are illustrated in greater detail in Fig.~\ref{fig:methodology}. The overall optimization process is summarized in Algorithm~\ref{alg:cando_main}.

\paragraph{In-Context Learning Initialization.}
\label{sec:method_init}
Given an input requirement $\mathcal{I}_\text{new}$ (Definition~\ref{def:instance}), the ICL Initializer agent produces an initial configuration $\mathbf{q}^{(0)}$ by conditioning layout synthesis on structurally similar examples retrieved from $\mathcal{D}^{\textup{dev}}$. The example retrieval process starts by embedding new input requirements $\mathcal{I}_\text{new}$ into a feature vector $\phi(\mathcal{I}_\text{new})\in\mathbb{R}^{d}$ covering four aspects: (i) per-class element counts and area coverages, (ii) access-point counts and their positions, (iii) element size statistics, and (iv) layout descriptors (area, perimeter, compactness). The most similar layouts to the query are retrieved with a weighted $k$NN search over the embedding space representing $\{\phi({\mathcal{I}})\mid \mathcal{I} \in \mathcal{D}^{\textup{dev}}\}$. The retrieved examples are serialized as input--output pairs (input: $\mathcal{I}$, output: $\mathbf{q}$) and provided as context to the agent, which is conditioned on $\mathcal{I}_\text{new}$. Next, $K_\text{init}$ candidate layouts are generated in parallel and ranked using a geometric hard-constraint satisfaction proxy:
\begin{equation}
  \sigma(\mathbf{q})
  \coloneqq \frac{1}{1 + V_{\mathrm{out}}(\mathbf{q}) + V_{\mathrm{ov}}(\mathbf{q})},~
  V_{\mathrm{out}}(\mathbf{q}) \coloneqq \frac{1}{|\mathcal{P}|}\sum_{i=1}^E |\Omega_i \setminus \mathcal{P}|,~
  V_{\mathrm{ov}}(\mathbf{q}) \coloneqq \frac{1}{|\mathcal{P}|}\sum_{i<j\leq E} |\Omega_i \cap \Omega_j|
  \label{eq:sigma}
\end{equation}
where $V_{\mathrm{out}}(\mathbf{q})$ and $V_{\mathrm{ov}}(\mathbf{q})$ are dimensionless out-of-premise and unique-pair overlap violations, normalized by premise area. Thus, $\sigma(\mathbf{q})=1$ exactly when the candidate satisfies containment and pairwise non-overlap. We rank the candidates by $\sigma(\cdot)$ and select the highest-scoring one. The result is returned as $\mathbf{q}^{(0)}$. Further details on specific feature definitions and the weighted $k$NN approach can be found in Appendix~\ref{app:icl-init}.

\paragraph{Verification.} It lies at the core of our proposed approach, providing \emph{reward signals} that guide the refinement of intermediate layouts. Given $K$ candidates in $\mathcal{Q}^{(t)}$, the Verifier evaluates them against the design manual $\mathcal{M}$ using dedicated deterministic tools and selects the best-scoring candidate $\mathbf{\bar q}^{(t)}$. The Verifier and benchmark evaluator intentionally share ground-truth-free design-manual checks (HC\textsubscript{g}, HC\textsubscript{f}, CO, PR, OR), but final scores are recomputed independently and the optimization never observes a held-out target layout. In particular, PL, MN, and PC compare against $\mathbf{q}^{\dagger}$ and are available only to the benchmark evaluator; hence, RE and AGG are not optimization signals. For richer diagnostics, the Verifier treats routing separately rather than integrating it into HCM, reports the five constraint aspects and CS, and aggregates fine-grained diagnostics (e.g., obstacles blocking a route) into $\mathcal{F}^{(t)}$ for the Historian and Planner.

\paragraph{History Management.}
The Historian agent maintains a running record of how the layout design evolves. After each step $t$, it computes the change in feedback and annotates it with a concise description of the spatial changes that led to it. A sliding window containing at most $W$ recent changes, $\mathcal{H}^{(t)} = \{\Delta \mathcal{F}^{(s)}\}_{s=\max(1,t-W+1)}^{t}$, is passed to the Planner, grounding its decisions in the evolution of feedback over time rather than only its current state.

\paragraph{Planner.} The Planner agent drafts its remediation plan along the same five evaluation categories used by the Verifier. Category-level plans are produced by the same agent, if the agent is capable of handling the full context (with a large backbone), or decomposed and processed in parallel by specialized planning sub-agents/workers, with task-specific tools, and then reconciled by a supervisor into a remediation plan $\Pi^{(t)}$ at each time step $t$ (as shown in Fig.~\ref{fig:methodology}). To resolve cross-category conflicts and prioritize repairs, the supervisor grounds its decisions in a \emph{multi-modal} input context defined by four complementary pieces of information: (i) the targeted diagnostic feedback from the specialist workers, (ii) the history $\mathcal{H}^{(t)}$, (iii) the geometric coordinate representation of the layout, and (iv) its visual rendering created by a simple plotting tool. By incorporating a subset or all of these signals, the supervisor formulates a coherent global intervention plan.

\paragraph{Generator.} This agent translates the remediation plan $\Pi^{(t)}$ into a set of $K$ revised candidate configurations $\mathcal{Q}^{(t)} = \{\mathbf{q}_k^{(t)}\}_{k=1}^{K}$ by applying a sequence of modifications to $\bar{\mathbf{q}}^{(t-1)} = \{(\mathbf{p}_i,\theta_i)\}_{i=1}^{E}$ (Eq.~\eqref{eq:placed}). The Generator adopts a strategy-diversified ensemble generation, with candidates either applying the plan conservatively through local repairs, prioritizing a single category or allowing larger layout-wide modifications. This ensemble design is important in practice because different layouts benefit from different remediation strategies. The strategy ensemble and their detailed definitions are discussed in detail in Appendix~\ref{app:generator-strategies}.

\begin{wrapfigure}[16]{r}{0.55\textwidth}
\centering \vspace{-0.4cm}
\begin{minipage}{0.95\linewidth}
    \begin{algorithm}[H]
        \fontsize{7pt}{10.5pt}\selectfont
        \caption{\texttt{CANDO}}
        \label{alg:cando_main}
        \begin{algorithmic}[1]
            \Require $\mathcal{I}$, $\mathcal{D}^{\mathrm{dev}}$, $N$, $B$, $K$, $W$, $T$
            \State $\mathcal{H}^{(0)} \gets \emptyset;~~t \gets 0$
            \State $(\mathbf{\bar q}^{(0)}, \mathcal{F}^{(0)}) \gets \textsc{Verifier}(\{\textsc{ICLInit}(\mathcal{I}, \mathcal{D}^{\mathrm{dev}})\})$
            \While {$t \leq T-B$} \label{line:rounds}
                \State $\{ \mathcal{H}_n^{(t)}, \mathcal{F}_n^{(t)}, \mathbf{q}_n^{(t)} \}_{n=1}^N \gets \textsc{Replicate}(\mathcal{H}^{(t)}, \mathcal{F}^{(t)}, \mathbf{\bar q}^{(t)})$\label{line:replicate}
                \For{$b = 0, \ldots, B-1$} \Comment{$\forall n \in [N]$, in parallel}\label{line:branching}
                    \State $\mathcal{H}_n^{(t+b+1)} \gets \textsc{Historian}(\mathcal{H}_n^{(t+b)},\,\mathcal{F}_n^{(t+b)},W)$\label{line:refinement-start}
                    \State $\Pi_n^{(t+b+1)} \gets \textsc{Planner}(\mathbf{q}_n^{(t+b)},\, \mathcal{F}_n^{(t+b)},\,\mathcal{H}_n^{(t+b+1)})$
                    \State $\mathcal{Q}_n^{(t+b+1)} \gets \textsc{Generator}(\mathbf{q}_n^{(t+b)},\,\Pi_n^{(t+b+1)}, K)$
                    \State $(\mathbf{\bar q}_n^{(t+b+1)}, \mathcal{F}_n^{(t+b+1)}) \gets \textsc{Verifier}(\mathcal{Q}_n^{(t+b+1)})$ \label{line:refinement-end}
                \EndFor
                \State $t \gets t+B$
                \State $(\mathbf{\bar q}^{(t)}, \mathcal{F}^{(t)}, \mathcal{H}^{(t)}) \gets \textsc{BoN}\!\left(\{\mathbf{\bar q}_n^{(t)}, \mathcal{F}_n^{(t)}, \mathcal{H}_n^{(t)}\}_{n=1}^{N}\right)$ \label{line:reduce-step}
            \EndWhile, ~ \Return $\mathbf{\bar q}^{(T)}$
        \end{algorithmic}
    \end{algorithm}
\end{minipage}
\end{wrapfigure}

\paragraph{Cooperative Refinement Loop.}
Starting from $\mathbf{\bar{q}}^{(0)}$, \texttt{CANDO} refines the layout through a parallel-branching loop that adapts CoDe's blockwise sampling scheme from diffusion models~\citep{singh2025code}, as summarized in Algorithm~\ref{alg:cando_main}. Until the budget of $T$ refinement steps is exhausted, in every iteration, \texttt{CANDO} spawns $N$ independent exploration branches from $\mathbf{\bar q}^{(t)}$ (Line~\ref{line:branching}). Every branch has its own local history ($\mathcal{H}_n^{(t)}$) and pursues its own trajectory of $B$ sequential steps. Each step corresponds to an execution of the history-aware planning-generation-verification sequence (Lines~\ref{line:refinement-start}--\ref{line:refinement-end}). At the end of each block ($B$ steps) a BoN reduce operation (Line~\ref{line:reduce-step}) selects the branch with the highest score (the new $\mathbf{\bar q}^{(t)}$), as well its corresponding history and feedback (the new $\mathcal{H}^{(t)}$ and $\mathcal{F}^{(t)}$, respectively), and passes it onto the next step where it will be replicated $N$ times to start the next block (Line~\ref{line:replicate}). The (block-wise) BoN approach enables the system to simultaneously explore structurally different refinements without committing prematurely to any single branch. Note that setting $B = T$ transforms CoDe back to BoN. The interested reader is referred to Appendix~\ref{app:methodology} for more elaborate discussions on the individual agents.
\section{Experiments}\vspace{-0.2cm}
\label{sec:experiments}
\textbf{Experimental Setup.} We assess \texttt{CANDO} on our proposed \textbf{ALPS-Bench} and three public benchmarks: PubLayNet~\citep{zhong2019publaynet}, RICO~\citep{deka2017rico}, and PKU-PosterLayout~\citep{hsu2023posterlayout}. For ALPS-Bench, the problem setting and corresponding evaluation metrics are defined in Sections~\ref{sec:formulation} and~\ref{sec:dataset}, respectively. For PubLayNet and RICO, we evaluate all tasks defined and adopted by prior art: type-conditioned generation (Gen-T), type/size-conditioned generation (Gen-TS), relational generation (Gen-R), refinement, and completion (latter three in Appendix~\ref{app:pubrico-additional-results}). A simplified version of \texttt{CANDO} is applied on the public benchmarks because not all agents are needed; its complete configuration is given in Appendix~\ref{app:simplified-cando}. For PubLayNet and RICO, we follow the evaluation criteria in the literature with the adjusted mIoU (as per \cite{layoutgan++}), alignment (``Align.''), overlap, and FID scores. For the latter three, we adopt the implementations of \citep{jiang2023layoutformer++}. For most of our experiments we opt for Qwen3.6~\citep{qwen36_35b_a3b} (Qwen/Qwen3.6-35B-A3B-FP8) as a backbone for our agents. This variant of Qwen is a lightweight, yet powerful open-source Mixture-of-Experts (MoE) \citep{jiang2024mixtral,fedus2022switch} model of $35$B total parameters and $3$B active parameters. For quantitative evaluations, we used NVIDIA H100 GPUs with $80$GB of VRAM. For our experiments with \textbf{ALPS-Bench}, we also provide experimental results with GPT-5.2 (API access via a cloud provider), a model with strong reasoning capabilities and orders of magnitude larger size.

\textbf{Results on ALPS-Bench.} Table~\ref{tab:cando_test_results} summarizes the performance of \texttt{CANDO} using our adaptation of CoDe's blockwise sampling procedure~\citep{singh2025code} with two backbone variants (Qwen and GPT-5.2). \textbf{ALPS-Bench} goes beyond standard generation tasks conditioned on type and size through an elaborate design manual. Therefore, few baselines apply directly to the newly defined tasks without considerable architectural modification and retraining. To compare against training-based approaches, we additionally adapt and train LayoutDM~\cite{inoue2023layoutdm} from scratch on \textbf{ALPS-Bench} and fine-tune Qwen3-VL-8B-Instruct~\cite{yang2025qwen3technicalreport} using LoRA-SFT~\cite{hu2021lora}. LayoutDM requires an adaptation of its native axis-aligned layout representation to accommodate the polygonal geometry and element orientations of \textbf{ALPS-Bench}, whereas Qwen-LoRA-ALPS generates the complete polygonal layout in a single pass conditioned on the rendered site and structured requirements. Adaptation details for both baselines are provided in Appendix~\ref{app:trained-baselines}. We further compare against the pioneering LLM-based approaches LayoutPrompter~\citep{lin2023layoutprompter} and LayoutCoT~\citep{shi2025layoutcot}, as well as our implementation of a Single-Agent using a strong backbone (GPT-5.2) and equipped with the evaluation scores and toolset devised for \texttt{CANDO}.

\begin{wrapfigure}{r}{0.5\textwidth}
  \vspace{-0.35cm}
  \centering
  \includegraphics[width=\linewidth]{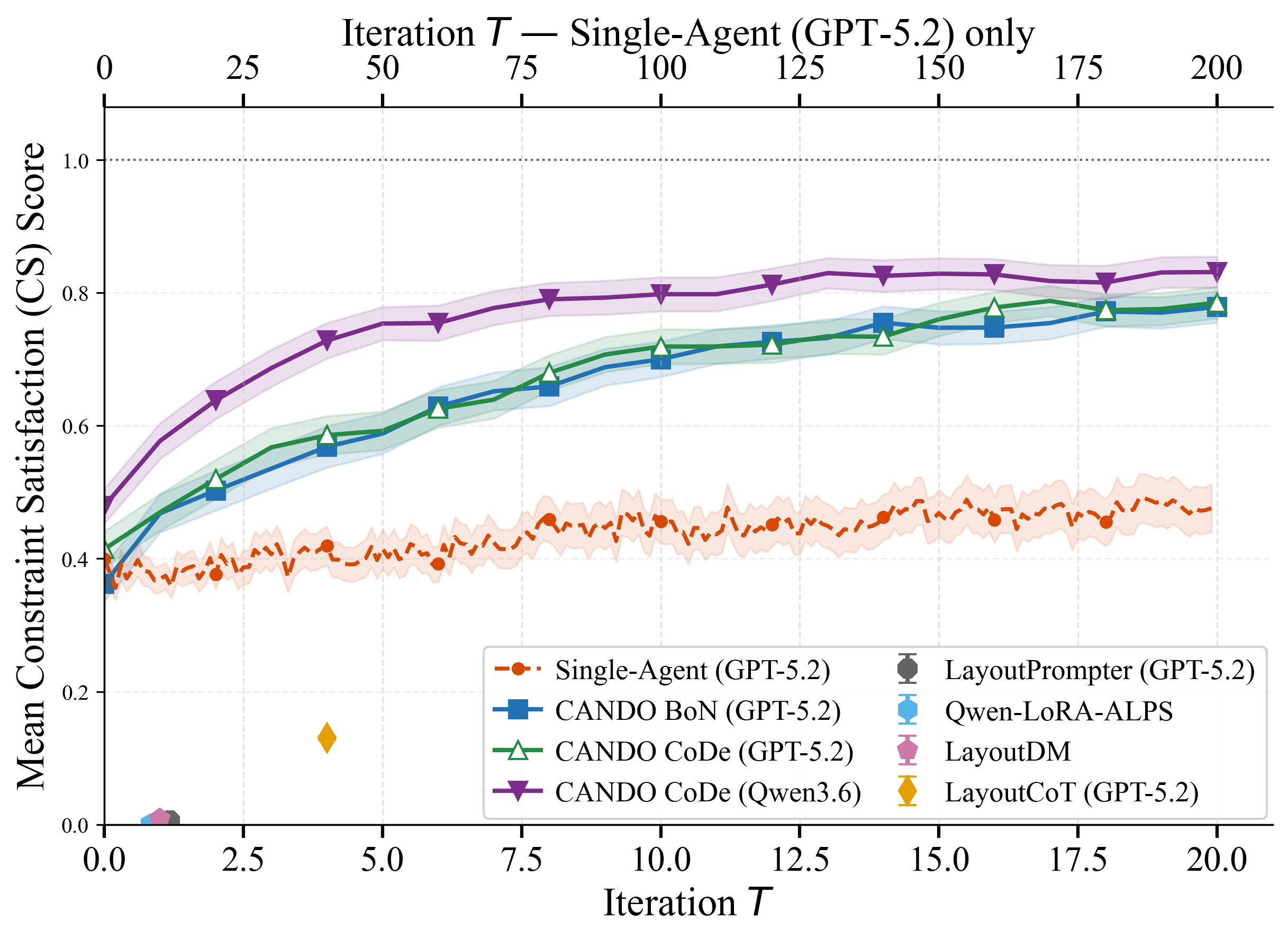}\vspace{-0.5cm}
  \caption{Mean CS Score vs. $T$ for Table~\ref{tab:cando_test_results}}
  \label{fig:cs-trajectory}\vspace{-0.3cm}
\end{wrapfigure} 

The summary statistics demonstrate that CANDO variants substantially outperform all considered baselines, including the two models trained specifically on \textbf{ALPS-Bench}, while requiring an order of magnitude smaller total number of refinement steps ($T$) than the Single-Agent baseline, with a time-evolution graph shown in Fig.~\ref{fig:cs-trajectory}. LayoutDM frequently violates geometric hard constraints because its bounding-box-based representation only approximates the true polygonal site geometry, while the single-pass Qwen-LoRA-ALPS can capture high-level compositional patterns but struggles with exact containment, non-overlap, and routing feasibility. These results suggest that dataset-specific training alone is insufficient for the hard geometric constraints of ALPS-Bench and further highlight the importance of iterative, tool-grounded verification. Additionally, paired, instance-level two-sided Wilcoxon tests with Holm correction and paired-bootstrap $95\%$ CIs over the $100$ test instances confirm that both \texttt{CANDO} variants significantly outperform all considered baselines on AGG, CS, and RE (all $p_{\mathrm{Holm}}\leq0.0059$), while the Qwen and GPT-5.2 variants are statistically comparable across all three metrics ($p_{\mathrm{Holm}}{=}1.000$). Full results are reported in Appendix~\ref{app:significance-testing}.

For the sake of fair comparison, we also provide time complexity of each approach averaged over the samples in $\mathcal{D}^{\text{test}}$. To measure time complexity, we assume a single LLM inference call to incur a fixed complexity of $\tau$. To provide a more complete picture, we compute the average number of iterations per method to reach compliance (on the samples achieving that). Considering a factor of $\times 4$ taking into account the number of sequential calls to each agent in \texttt{CANDO}, we arrive at the summary provided in the table. The results show that \texttt{CANDO} requires the same order of complexity (half or even less, to be precise). This underscores that despite the increased operational complexity per step, the parallelized nature of \texttt{CANDO} yields a competitive overall complexity that scales more favorably than the extensive sequential Single-Agent counterpart.

\begin{table}[t]
\centering
\caption{Quantitative results of layout generation models evaluated on the \textbf{ALPS-Bench} test set ($\mathcal{D}^{\textup{test}}$). $^\dagger$ denotes methods trained or fine-tuned on ALPS-Bench; all other methods are training-free with respect to ALPS-Bench. Symbols indicate the generation paradigm: $\blacklozenge$ diffusion-based, $\blacktriangle$ single-LLM/MLLM, and $\bullet$ agentic.}
\label{tab:cando_test_results}
\resizebox{\linewidth}{!}{
    \begin{tabular}{l ccc c c c c}
        \toprule
        \multirow{2}{*}{\textbf{Method}} & \multicolumn{3}{c}{\textbf{Summary Statistics}} & \multirow{2}{*}{$T$} & \multirow{2}{*}{\textbf{\shortstack{Avg. Iterations \\ to Comply}}} & \multirow{2}{*}{\textbf{\shortstack{Time Complexity}}} & \multirow{2}{*}{\textbf{Compliance (\%)}} \\
        \cmidrule{2-4}
        & \textbf{AGG}$\uparrow$ & \textbf{CS}$\uparrow$ & \textbf{RE}$\uparrow$ & & & & \\
        \midrule
        $\blacktriangle$ Qwen-LoRA-ALPS$^\dagger$ & $0.07_{\pm 0.12}$ & $0.01_{\pm 0.08}$ & $0.13_{\pm 0.22}$ & $1$ & -- & $\tau$ & -- \\
        $\blacklozenge$ LayoutDM$^\dagger$ \cite{inoue2023layoutdm} & $0.13_{\pm 0.13}$ & $0.01_{\pm 0.03}$ & $0.24_{\pm 0.25}$ & -- & -- & -- & -- \\
        $\blacktriangle$ LayoutPrompter (GPT-5.2)~\citep{lin2023layoutprompter} & $0.17_{\pm 0.11}$ & $0.01_{\pm 0.06}$ & $0.33_{\pm 0.21}$ & $1$ & -- & $\tau$ & -- \\
        $\blacktriangle$ LayoutCoT (GPT-5.2)~\citep{shi2025layoutcot} & $0.19_{\pm 0.20}$ & $0.13_{\pm 0.24}$ & $0.25_{\pm 0.24}$ & $4$ & -- & $4\tau$ & -- \\
        $\bullet$ Single-Agent (GPT-5.2) & $0.40_{\pm 0.27}$ & $0.47_{\pm 0.37}$ & $0.32_{\pm 0.22}$ & $200$ & $80.6$ & $80.6\tau$ & $22.4$ \\
        \midrule
        $\bullet$ \texttt{CANDO} (BoN/GPT-5.2) & $0.59_{\pm 0.25}$ & $0.77_{\pm 0.36}$ & $0.41_{\pm 0.17}$ & $20$ & $8.55$ & $8.55 \times 4\tau = 34.2\tau$ & $60.6$ \\
        $\bullet$ \texttt{CANDO} (CoDe/GPT-5.2) & $0.60_{\pm 0.25}$ & $0.78_{\pm 0.35}$ & $\mathbf{0.42_{\pm 0.17}}$ & $20$ & $8.81$ & $8.81 \times 4\tau = 35.24\tau$ & $63.6$ \\
        $\bullet$ \texttt{CANDO} (CoDe/Qwen3.6) & $\mathbf{0.62_{\pm 0.19}}$ & $\mathbf{0.83_{\pm 0.25}}$ & $0.41_{\pm 0.16}$ & $20$ & $\mathbf{5.09}$ & $5.09 \times 4\tau = \mathbf{20.36\tau}$ & $\mathbf{74.0}$ \\
        \bottomrule
    \end{tabular}
}
\end{table}
\begin{table}[t]
\centering
\caption{Quantitative results on PubLayNet and RICO datasets. Extended version in Appendix~\ref{app:pubrico-additional-results}.}
\label{tab:results_publaynet_rico_main}
\resizebox{0.9\linewidth}{!}{
    \begin{tabular}{ll cccc c cccc}
        \toprule
        & & \multicolumn{4}{c}{\cellcolor{ricoColor}\textbf{RICO~\citep{deka2017rico}}} && \multicolumn{4}{c}{\cellcolor{publayColor}\textbf{PubLayNet~\citep{zhong2019publaynet}}} \\
        \cmidrule{3-6} \cmidrule{8-11}
        \textbf{Task} & \textbf{Method} & \textbf{mIoU}$\uparrow$ & \textbf{FID}$\downarrow$ & \textbf{Align}$\downarrow$ & \textbf{Over}$\downarrow$ && \textbf{mIoU}$\uparrow$ & \textbf{FID}$\downarrow$ & \textbf{Align}$\downarrow$ & \textbf{Over}$\downarrow$ \\
            \midrule
        \multirow{6}{*}{\textbf{Gen-T}}
        & LayoutGAN++~\citep{layoutgan++} & 0.298 & 5.954 & 0.261 & 0.620 && 0.297 & 14.875 & 0.124 & 0.148 \\
        & BLT~\citep{kong2022blt} & 0.216 & 25.633 & 0.150 & 0.983 && 0.140 & 38.684 & 0.036 & 0.196 \\
        & LayoutFormer++~\citep{jiang2023layoutformer++} & 0.432 & \textbf{1.096} & 0.230 & 0.530 && 0.348 & 8.411 & 0.020 & 0.008 \\
        & LayoutPrompter~\citep{lin2023layoutprompter} & 0.604 & 2.134 & 0.002 & 0.029 && 0.619 & \underline{1.648} & 0.001 & 0.004 \\
        & LayoutCoT~\citep{shi2025layoutcot} & \underline{0.719} & 2.189 & \underline{0.001} & \underline{0.022} && \underline{0.772} & \textbf{1.227} & \textbf{0.000} & \underline{0.002} \\
        \cmidrule{2-11}
        & \cellcolor{candoBlue!10}\texttt{CANDO} (ours) & \cellcolor{candoBlue!10}\textbf{0.866} & \cellcolor{candoBlue!10}\underline{2.101} & \cellcolor{candoBlue!10}\textbf{0.000} & \cellcolor{candoBlue!10}\textbf{0.005} && \cellcolor{candoBlue!10}\textbf{0.893} & \cellcolor{candoBlue!10}3.455 & \cellcolor{candoBlue!10}\textbf{0.000} & \cellcolor{candoBlue!10}\textbf{0.000} \\
        \midrule
        \multirow{5}{*}{\textbf{Gen-TS}}
        & BLT~\citep{kong2022blt} & 0.604 & \underline{0.951} & 0.181 & 0.660 && 0.428 & 7.914 & 0.021 & 0.419 \\
        & LayoutFormer++~\citep{jiang2023layoutformer++} & 0.620 & \textbf{0.757} & 0.202 & 0.542 && 0.471 & \textbf{0.720} & 0.024 & 0.037 \\
        & LayoutPrompter~\citep{lin2023layoutprompter} & 0.707 & 3.072 & 0.002 & \underline{0.051} && 0.727 & 1.863 & 0.001 & 0.076 \\
        & LayoutCoT~\citep{shi2025layoutcot} & \underline{0.795} & 3.506 & \textbf{0.000} & \textbf{0.031} && \underline{0.847} & \underline{1.430} & \textbf{0.000} & \underline{0.004} \\
        \cmidrule{2-11}
        & \cellcolor{candoBlue!10}\texttt{CANDO} (ours) & \cellcolor{candoBlue!10}\textbf{0.879} & \cellcolor{candoBlue!10}1.371 & \cellcolor{candoBlue!10}\textbf{0.000} & \cellcolor{candoBlue!10}0.052 && \cellcolor{candoBlue!10}\textbf{0.881} & \cellcolor{candoBlue!10}2.023 & \cellcolor{candoBlue!10}\textbf{0.000} & \cellcolor{candoBlue!10}\textbf{0.000} \\
        \bottomrule
    \end{tabular}
}
\end{table}

\textbf{Results on PubLayNet and RICO.} To further assess the efficacy of \texttt{CANDO} beyond the proposed \textbf{ALPS-Bench}, we extended our evaluations against state-of-the-art layout generation models on two widely adopted benchmarks: PubLayNet and RICO. Table~\ref{tab:results_publaynet_rico_main} summarizes the performance for type-conditioned (Gen-T) and type/size-conditioned (Gen-TS) layout generation tasks. \texttt{CANDO} achieves the highest mIoU on both settings by a substantial margin, and perfect alignment scores across both benchmarks and tasks. This is noteworthy given that here \texttt{CANDO} runs on a comparatively lightweight open-source backbone Qwen with $35$B parameters compared to GPT-4 used in LayoutCoT~\citep{shi2025layoutcot} and the most recent (and the best) reported results for LayoutPrompter~\citep{lin2023layoutprompter}. Across all five RICO tasks, \texttt{CANDO} also improves overlap over LayoutCoT in four cases, with Gen-TS as the exception (Appendix~\ref{app:pubrico-additional-results}). These alignment and overlap results reflect \texttt{CANDO}'s verification and tool-calling strategies. By offloading axis-alignment and overlap checks to dedicated geometric tools, rather than relying on the (M)LLM to reason about coordinates, \texttt{CANDO} handles such violations precisely.

The FID scores here present a more nuanced picture. Despite leading on mIoU, \texttt{CANDO} does not consistently achieve competitive FID, ranking second or last depending on the setting. This inconsistency is not incidental but rather symptomatic of the broader limitation of distributional evaluation we discussed in Sections~\ref{sec:intro} and~\ref{sec:dataset}. We argue that a method that produces geometrically precise layouts that satisfy desired constraints and requirements yet happens to deviate from the distribution of the training data will incur a high FID penalty, even if its outputs are arguably of higher design quality. Methods such as LayoutFormer++, which are trained explicitly against the prior distribution, are inherently better at minimizing FID, even though as the mIoU results suggest, at the cost of spatial accuracy. Notably, this is known in the context of reward-aligned foundation models (analogous to meeting design requirements through a verification scheme in our setting) where more innovative samples can be generated at the cost of larger KL divergence from prior data \citep{jayasumana2024rethinking, singh2025code}.

\begin{wraptable}[9]{R}{0.60\linewidth}
\vspace{-0.45cm}
\centering
\caption{Quantitative results on PKU-PosterLayout dataset. Additional tasks are reported in Appendix~\ref{app:pku-additional-results}.}
\label{tab:results_pku_main}
\resizebox{\linewidth}{!}{
    \begin{tabular}{ll ccccc}
        \toprule
        \textbf{Task} & \textbf{Method} &
        \textbf{Occ}$\downarrow$ &
        \textbf{Rea}$\downarrow$ &
        \textbf{Und$_S$}$\uparrow$ &
        \textbf{Ove}$\downarrow$ &
        \textbf{FID}$\downarrow$ \\
        \midrule
        \multirow{3}{*}{\textbf{Gen-T}}
        & RALF~\cite{horita2024retrieval}       & \underline{0.136} & \textbf{0.0127} & 0.890 & 0.0091 & \textbf{2.36} \\
        & UniLayDiff~\cite{liu2026unilaydiff} & \textbf{0.122} & \underline{0.0129} & \underline{0.963} & \underline{0.0018} & \underline{3.09} \\
        & \cellcolor{candoBlue!10}\texttt{CANDO} (ours)
                      & \cellcolor{candoBlue!10}0.144
                      & \cellcolor{candoBlue!10}0.0196
                      & \cellcolor{candoBlue!10}\textbf{1.000}
                      & \cellcolor{candoBlue!10}\textbf{0.0009}
                      & \cellcolor{candoBlue!10}9.83 \\
        \midrule
        \multirow{3}{*}{\textbf{Gen-TS}}
        & RALF~\cite{horita2024retrieval} & \underline{0.138} & \textbf{0.0123} & 0.876 & 0.0113 & \textbf{0.61} \\
        & UniLayDiff~\cite{liu2026unilaydiff} & \textbf{0.128} & \underline{0.0132} & \underline{0.918} & \underline{0.0046} & \underline{0.66} \\
        & \cellcolor{candoBlue!10}\texttt{CANDO} (ours)
                      & \cellcolor{candoBlue!10}0.145
                      & \cellcolor{candoBlue!10}0.0197
                      & \cellcolor{candoBlue!10}\textbf{0.990}
                      & \cellcolor{candoBlue!10}\textbf{0.0000}
                      & \cellcolor{candoBlue!10}4.47 \\
        \bottomrule
    \end{tabular}
}
\end{wraptable}
\textbf{Results on PKU-PosterLayout.}
To further assess the generalizability of \texttt{CANDO}, we additionally adapt and evaluate it on PKU-PosterLayout, considering its controllable generation tasks. Table~\ref{tab:results_pku_main} reports results for type-conditioned (Gen-T) and type/size-conditioned (Gen-TS) layout generation tasks. Results for relationship-conditioned generation, completion, and refinement are deferred to Appendix~\ref{app:pku-additional-results}. Across both tasks, \texttt{CANDO} exhibits particularly strong performance on metrics directly associated with explicit spatial constraint satisfaction. Specifically, it attains the best Strict Underlay scores ($1.000$ and $0.990$) while simultaneously reducing overlap to $0.0009$ and $0.0$, respectively, outperforming both RALF~\cite{horita2024retrieval} and UniLayDiff~\cite{liu2026unilaydiff} on these metrics. Conversely, \texttt{CANDO} obtains higher FID scores and does not consistently improve occlusion or readability. This behavior is consistent with the observations on PubLayNet and RICO: \texttt{CANDO}'s verification-driven generation procedure explicitly optimizes compliance with geometric and relational requirements rather than reproducing the empirical layout distribution of the dataset, as measured by distribution-sensitive metrics such as FID.

\begin{table}[t]
\centering
\caption{Component ablation of \texttt{CANDO}. Regression rate denotes the percentage of refinement iterations for which the verification score decreases after applying the proposed changes.}
\label{tab:component_ablation}
\resizebox{\linewidth}{!}{
    \begin{tabular}{l ccc ccc}
        \toprule
        \textbf{Variant} & \textbf{AGG}$\uparrow$ & \textbf{CS}$\uparrow$ & \textbf{RE}$\uparrow$ & \textbf{\shortstack{Verifier-Compliant\\Layouts}} & \textbf{\shortstack{Mean\\Iterations}} & \textbf{\shortstack{Regression\\Rate}} \\
        \midrule
        \texttt{CANDO} (full) & $0.62_{\pm 0.19}$ & $0.83_{\pm 0.25}$ & $0.41_{\pm 0.16}$ & 74\% & 13.28 & 7.0\% \\
        \midrule
        w/o CoDe & $0.35_{\pm 0.30}$ & $0.38_{\pm 0.45}$ & $0.32_{\pm 0.23}$ & 44\% & 16.75 & 6.3\% \\
        w/o Historian & $0.61_{\pm 0.23}$ & $0.80_{\pm 0.34}$ & $0.42_{\pm 0.16}$ & 59\% & 13.35 & 6.7\% \\
        w/o ICL & $0.57_{\pm 0.26}$ & $0.75_{\pm 0.37}$ & $0.40_{\pm 0.19}$ & 53\% & 14.39 & 5.8\% \\
        w/o Planner & $0.62_{\pm 0.22}$ & $0.82_{\pm 0.32}$ & $0.43_{\pm 0.15}$ & 51\% & 14.63 & 6.7\% \\
        w/o Strategies & $0.61_{\pm 0.23}$ & $0.80_{\pm 0.32}$ & $0.42_{\pm 0.16}$ & 53\% & 14.60 & 18.4\% \\
        w/o Tool-grounded Verifier & $0.27_{\pm 0.23}$ & $0.28_{\pm 0.27}$ & $0.25_{\pm 0.24}$ & 95\%$^{*}$ & 4.63 & 19.6\% \\
        w/o Visual Rendering & $0.61_{\pm 0.24}$ & $0.80_{\pm 0.35}$ & $0.42_{\pm 0.16}$ & 61\% & 13.35 & 7.3\% \\
        \bottomrule
    \end{tabular}
}
{\footnotesize $^{*}$Without tool access, the Verifier acts solely as an LLM-as-a-Judge, inflating its scores.}
\end{table}

\textbf{Component Ablation.} Table~\ref{tab:component_ablation} assesses the contribution of the main architectural components of \texttt{CANDO}. We define the \emph{regression rate} as the percentage of refinement iterations in which the verification score decreases after applying the proposed changes, capturing how often a refinement step inadvertently introduces or worsens constraint violations. CoDe and tool-grounded verification have the most pronounced impact. Replacing CoDe with a single refinement trajectory reduces AGG from $0.62$ to $0.35$, CS from $0.83$ to $0.38$, and verifier-compliant layouts from $74\%$ to $44\%$, while increasing the mean number of iterations from $13.28$ to $16.75$. This demonstrates the importance of parallel exploration for escaping difficult constraint violations rather than committing to a single refinement trajectory. Tool-grounded verification is similarly critical. Without deterministic verification tools, the LLM-based Verifier declares $95\%$ of layouts compliant and terminates after only $4.63$ iterations on average, despite the independent evaluation yielding only $0.27$ AGG and $0.28$ CS. The corresponding regression rate also increases from $7.0\%$ to $19.6\%$. This discrepancy highlights that an LLM-as-a-Judge can substantially overestimate constraint satisfaction and corroborates the importance of grounding feedback in deterministic tools.

Strategy diversification primarily affects the stability of refinement. Removing the alternative refinement strategies leaves the average scores comparatively unchanged, but increases the regression rate from $7.0\%$ to $18.4\%$ and reduces compliance from $74\%$ to $53\%$. Thus, a single generic repair strategy is substantially more likely to resolve one violation while introducing another. The distribution of winning strategies, reported in Appendix~\ref{app:component-ablation}, further shows that no individual strategy dominates the refinement process. ICL provides the next most pronounced benefit, with its removal reducing AGG from $0.62$ to $0.57$, CS from $0.83$ to $0.75$, and compliance from $74\%$ to $53\%$, while requiring more refinement iterations. The remaining components primarily improve robustness rather than mean performance. Notably, removing the Planner leaves mean AGG nearly unchanged ($0.62$) but reduces compliance from $74\%$ to $51\%$, suggesting that direct Verifier feedback can often repair individual violations but is less reliable at jointly reconciling all constraints.

\textbf{Hyperparameter Tuning.} Detailed hyperparameter ablations for \textbf{ALPS-Bench} are reported in Appendix~\ref{app:hyperparameter-ablation}. Based on a development-set sweep using Qwen for cost efficiency, we select $N{=}4$, $B{=}10$, $K{=}6$, and $W{=}6$, which achieve $0.76$ CS and $0.59$ AGG. We retain this configuration across backbone variants and increase the refinement budget to $T{=}20$ for the reported ALPS-Bench experiments. Appendix~\ref{app:simplified-cando} details the fixed configurations used on all public benchmarks.
\section{Concluding Remarks}\vspace{-0.2cm}
\label{sec:conc}
We introduced \textbf{ALPS-Bench}, a benchmark of $1,000$ professionally annotated facility layouts with an instance-specific scoring protocol that measures actual constraint satisfaction, and \texttt{CANDO}, a training-free cooperative agentic framework that refines layouts through a verification-grounded loop. \texttt{CANDO} establishes a strong baseline on \textbf{ALPS-Bench} and achieves state-of-the-art mIoU and alignment on PubLayNet and RICO with a comparatively lightweight backbone. On PKU-PosterLayout, it obtains the lowest overlap across all five tasks and the highest strict-underlay score in four of five tasks. These results corroborate that cooperative agentic design is a broadly effective recipe for constraint-aware layout synthesis. We hope \textbf{ALPS-Bench} shifts the focus of layout generation research beyond the current regime and toward the operational complexity of real-world design, and that \texttt{CANDO} motivates further work in this direction, positioning cooperative agentic design with tool-grounded verification as a strong recipe for constraint-aware synthesis.

\textbf{Limitations.} The proposed \textbf{ALPS-Bench} is built from outdoor service facility layouts, and transferring \texttt{CANDO} to a new domain requires a corresponding design manual and, if ICL is used, a development corpus. The parallel-branching strategy also incurs more (M)LLM calls than single-agent baselines, with a clear performance-cost tradeoff (Table~\ref{tab:hyperparameter_tuning}). Finally, \texttt{CANDO} does not optimize for distributional similarity, yielding strong mIoU but weaker FID on PubLayNet, RICO, and PKU.

\begin{ack}
\textbf{Acknowledgments.} The authors thank Anuj Singh (formerly Shell, now TU Delft) for the insightful technical ideations and help with data collection, Stylianos Poulakakis-Daktylidis for his assistance with data curation, as well as Prof. Hamed Pirsiavash (UC Davis) for early insightful discussions. 

\textbf{Funding in direct support of this work.} This work was funded by Shell.

\textbf{Competing interests.} The authors declare no competing interests.
\end{ack}

\bibliographystyle{plainnat}
\bibliography{references}

\newpage
\appendix
\section{Broader Impact}
\label{app:broader-impact}
We do hope that the \textbf{ALPS-Bench} shifts the focus of layout generation research beyond the current regime and toward the operational complexity of real-world design, and that \texttt{CANDO} motivates further work in this direction, positioning cooperative agentic design with tool-grounded verification as a strong recipe for constraint-aware synthesis.

The proposed framework operates on geometric layout representations derived from professional CAD drawings and does not process personal data, photographs, or any content tied to individuals. The introduced \textbf{ALPS-Bench} dataset consists solely of anonymized site geometries and does not contain identifying information about facility operators or locations. We are not aware of a direct path to harmful applications. The framework generates two-dimensional element placements for operational sites and has no generative capability in domains such as media synthesis or surveillance.

A secondary consideration is automation-induced risk in safety-critical design. While \texttt{CANDO} is intended as a decision-support tool rather than a replacement for human review, its outputs could, in principle, be adopted with insufficient expert oversight. We emphasize that the routing feasibility checks implemented in the Verifier and Evaluation Protocol are conservative by design (inflated obstacle margins, strict no-backtracking rules), and that final layouts should always be reviewed by a qualified engineer before deployment.

\section{Limitations}
\label{app:limitations}

\paragraph{Domain scope of \textbf{ALPS-Bench}.} The benchmark is constructed from outdoor automotive service facility layouts. Although the problem formulation in Section~\ref{sec:formulation} is domain-agnostic, the design manual $\mathcal{M}$, the routing scenarios in Appendix~\ref{app:routing}, and the 14-class element vocabulary are all specific to this setting. Transferring \texttt{CANDO} to a new domain requires authoring a corresponding design manual and, if ICL initialization is used, a development corpus of annotated examples.

\paragraph{Computational Cost.} The parallel-branching strategy of \texttt{CANDO} with $N$, $B$, and $K$ incurs more LLM calls per layout than Single-Agent or LayoutPrompter baselines. The ablation in Table~\ref{tab:hyperparameter_tuning} shows a clear performance--cost tradeoff: the highest-scoring configuration is also the most expensive. Deployment in resource-constrained settings may require reducing $N$, $B$, or $K$ at some cost to layout quality.

\paragraph{FID and Distributional Evaluation.} \texttt{CANDO} does not optimize for distributional similarity to training data. As discussed in Section~\ref{sec:experiments}, this leads to state-of-the-art performance in mIoU and alignment scores on PubLayNet and RICO, but inconsistent FID rankings. Researchers whose evaluation protocol weights FID heavily should treat these results with the caveats noted in Section~\ref{sec:intro}.

\clearpage
\section{Notation}\label{app:notation}

Table~\ref{tab:notation} summarizes all mathematical symbols used throughout the paper, grouped by the section in which they are introduced.

\begin{longtable}{@{} l p{9.5cm} @{}}
\caption{Summary of mathematical notation used throughout the paper.}\label{tab:notation}\\
    \toprule
    \textbf{Symbol} & \textbf{Description} \\
    \midrule
    \endfirsthead
    \multicolumn{2}{@{}l}{\small\itshape (continued from previous page)}\\[2pt]
    \toprule
    \textbf{Symbol} & \textbf{Description} \\
    \midrule
    \endhead
    \midrule
    \multicolumn{2}{@{}r}{\small\itshape (continued on next page)}\\
    \endfoot
    \bottomrule
    \endlastfoot
    \multicolumn{2}{@{}l}{Layout domain \textnormal{(Section~\ref{sec:formulation})}} \\[2pt]
    \midrule
    $\mathcal{C}$ & Vocabulary of semantic element classes. \\
    $E$ & Number of elements to be placed. \\
    $\mathcal{A}$ & Set of designated entrance and exit access regions. \\
    $\mathcal{E} = \{(c_i,\,G_i)\}_{i=1}^{E}$ & Full element specification (fixed input). \\
    $G_i \subset \mathbb{R}^2$ & Canonical (fixed) shape of element $i$, centered at the origin. \\
    $\mathcal{P} \subset \mathbb{R}^2$ & Layout Premise -- the outer boundary polygon. \\
    $\mathcal{Q}$ & Space of all possible configurations. \\
    $\mathcal{Q}_{\mathrm{valid}} \subset \mathcal{Q}$ & Subset of configurations satisfying all hard constraints. \\
    $R(\theta_i)$ & $2{\times}2$ rotation matrix for angle $\theta_i$. \\
    $c_i \in \mathcal{C}$ & Semantic class of element $i$. \\
    $e_i = (c_i,\,G_i)$ & Element $i$, with semantic class $c_i$ and canonical shape $G_i$. \\
    $\mathbf{p}_i = (x_i,\,y_i) \in \mathbb{R}^2$ & Centroid of placed element $i$ (optimization variable). \\
    $\mathbf{q} = \{(\mathbf{p}_i,\theta_i)\}_{i=1}^{E}$ & Elements configuration -- the full set of placement parameters. \\
    $\Omega_i(\mathbf{p}_i,\theta_i) = R(\theta_i)\,G_i + \mathbf{p}_i$ & Placed geometry of element $i$ after rotation and translation. \\
    $\theta_i \in [0,2\pi)$ & Orientation angle of element $i$ (optimization variable). \\
    \addlinespace[4pt]
    \multicolumn{2}{@{}l}{\textit{Constraints \& objective \textnormal{(Section~\ref{sec:formulation})}}} \\[2pt]
    \midrule
    $\mathcal{HC}$ & Hard constraints (containment, non-overlap, routing feasibility). \\
    $\mathcal{I} = (\mathcal{P},\,\mathcal{A},\,\mathcal{E},\,\mathcal{M})$ & Problem instance. \\
    $\mathcal{J}(\mathbf{q}) = \sum_{s=1}^{S} w_s\,f_s(\mathbf{q})$ & Weighted soft-constraint objective. \\
    $\mathcal{M} = \{\mathcal{HC},\,\mathcal{SC}\}$ & Design manual, partitioned into hard and soft constraints. \\
    $\mathcal{O}_r(\mathbf{q})$ & Set of ground-plane elements that block traversal in routing scenario $r$. \\
    $\mathcal{R}$ & Set of predefined routing scenarios. \\
    $\mathcal{F}_r(\mathbf{q})$ & Scenario-specific free space $\mathcal{P}\setminus\bigcup_{i\in\mathcal{O}_r(\mathbf{q})}\Omega_i$. \\
    $S = |\mathcal{SC}|$ & Number of soft constraints. \\
    $\mathcal{SC}$ & Soft constraints (compositionality, orientation, proximity). \\
    $f_s(\mathbf{q}) \in [0,1]$ & Scoring function measuring compliance with soft constraint $s$. \\
    $\mathbf{q}^{\star}$ & Optimal (highest-scoring) configuration. \\
    $w_s$ & Weight for soft constraint $s$;\ $\sum_{s=1}^{S}w_s = 1$. \\
    \addlinespace[4pt]
    \multicolumn{2}{@{}l}{\textit{Dataset \& evaluation metrics \textnormal{(Section~\ref{sec:dataset})}}} \\[2pt]
    \midrule
    $\mathrm{AGG}$ & Aggregate score $=(\mathrm{CS}+\mathrm{RE})/2$. \\
    CO  & Compositionality score $\in [0,1]$. \\
    $\mathrm{CS}$ & Constraint satisfaction score defined in Eq.~\eqref{eq:agg_metrics}. \\
    $\mathcal{D}^{\textup{dev}}$ & Development set (900 annotated instances). \\
    $\mathcal{D}^{\textup{test}}$ & Held-out test set (100 instances). \\
    $\mathrm{HC}_{f}$ & Flow score: fraction of routing scenarios with a feasible path. \\
    $\mathrm{HC}_{g}$ & Geometric hard-constraint score penalising out-of-boundary and overlap areas. \\
    $\mathrm{HCM}$ & Hard constraints metric: $0$ if elements are missing or incorrectly sized; otherwise $\alpha_{\mathrm{g}}\mathrm{HC}_{g}+\alpha_{\mathrm{f}}\mathrm{HC}_{f}$. \\
    $\alpha_{\mathrm{g}},\alpha_{\mathrm{f}}$ & Geometric-validity and flow weights in $\mathrm{HCM}$, respectively; $(\alpha_{\mathrm{g}},\alpha_{\mathrm{f}})=(0.2,0.8)$. \\
    MN  & Maneuverability sub-score $\in [0,1]$ ($= \overline{n}_{\mathrm{gt}} / (\overline{n}_{\mathrm{gt}} + \overline{n}_{\mathrm{gen}})$). \\
    OR  & Orientation score $\in [0,1]$. \\
    PC  & Path clearance sub-score $\in [0,1]$ ($= \overline{c}_{\mathrm{gen}} / (\overline{c}_{\mathrm{gt}} + \overline{c}_{\mathrm{gen}})$). \\
    PL  & Path length sub-score $\in [0,1]$ ($= \overline{\ell}_{\mathrm{gt}} / (\overline{\ell}_{\mathrm{gt}} + \overline{\ell}_{\mathrm{gen}})$). \\
    PR  & Proximity score $\in [0,1]$. \\
    $\mathrm{RE}$ & Routing Efficiency $= ({\rm PL} + {\rm MN} + {\rm PC})/3$. \\
    $\mathrm{SCM}$ & Soft constraints metric $= \alpha_{\mathrm{p}}\mathrm{PR}+\alpha_{\mathrm{c}}\mathrm{CO}+\alpha_{\mathrm{o}}\mathrm{OR}$. \\
    $\alpha_{\mathrm{p}},\alpha_{\mathrm{c}},\alpha_{\mathrm{o}}$ & Proximity, compositionality, and orientation weights in $\mathrm{SCM}$, respectively; $(\alpha_{\mathrm{p}},\alpha_{\mathrm{c}},\alpha_{\mathrm{o}})=(0.6,0.3,0.1)$. \\
    $\alpha$ & Penalty cap for invalid layouts in $\mathrm{CS}$; $\alpha=0.5$. \\
    $\overline{c}$ & Mean minimum path clearance over routing scenarios. \\
    $\overline{\ell}$ & Mean path length over routing scenarios. \\
    $\overline{n}$ & Mean number of turns over routing scenarios. \\
    $(\cdot)_{\mathrm{gt}},(\cdot)_{\mathrm{gen}}$ & Quantity measured for the ground-truth or generated layout, respectively. \\
    $\mathbf{q}^{\dagger}$ & Ground-truth configuration. \\
    $\epsilon$ & Scenario-dependent obstacle-inflation safety margin (Voronoi planning). \\
    \addlinespace[4pt]
    \multicolumn{2}{@{}l}{\textit{\texttt{CANDO} framework \textnormal{(Section~\ref{sec:methodology})}}} \\[2pt]
    \midrule
    $B$ & Block size: number of sequential Planner--Generator--Verifier steps per branch. \\
    $B_t$ & Number of steps in the current block, $\min(B,T-t)$. \\
    $b$ & Within-block step offset, $b\in\{0,\ldots,B_t-1\}$. \\
    $\mathcal{F}^{(t)},\mathcal{F}_n^{(t)}$ & Structured feedback report for the selected global state or branch $n$ at step $t$. \\
    $\Delta\mathcal{F}^{(t)}$ & Change in feedback at step $t$, annotated with the spatial changes that produced it. \\
    $\mathcal{H}^{(t)},\mathcal{H}_n^{(t)}$ & Sliding-window history for the selected global state or branch $n$ at step $t$. \\
    $\mathcal{I}_{\mathrm{new}}$ & Query (target) problem instance presented to \texttt{CANDO}. \\
    $K$ & Number of candidate configurations produced by the Generator per step. \\
    $K_{\mathrm{init}}$ & Number of candidate layouts generated by the ICL Initializer. \\
    $N$ & Number of parallel exploration branches per outer round. \\
    $\mathcal{Q}^{(t)},\mathcal{Q}_n^{(t)}$ & Set of $K$ candidate configurations generated globally or for branch $n$ at step $t$. \\
    $T$ & Maximum total number of optimization steps. \\
    $V_{\mathrm{out}}(\mathbf{q})$ & Out-of-premise area summed over elements and normalized by $|\mathcal{P}|$. \\
    $V_{\mathrm{ov}}(\mathbf{q})$ & Unique-pair overlap area summed over $i<j$ and normalized by $|\mathcal{P}|$. \\
    $W$ & History sliding window size. \\
    $\mathbf{q}^{(0)}$ & Initial seed configuration produced by the ICL Initializer. \\
    $\bar{\mathbf{q}}^{(t)}$ & Best-scoring candidate configuration selected at step $t$. \\
    $\mathbf{q}_n^{(t)}$ & Configuration of branch $n$ at inner step $t$. \\
    $t$ & Optimization step index. \\
    $\Pi^{(t)},\Pi_n^{(t)}$ & Remediation plan produced globally or for branch $n$ at step $t$. \\
    $\sigma(\mathbf{q})$ & Hard-constraint proxy feasibility score (Eq.~\ref{eq:sigma}). \\
    $\tau$ & Time complexity unit corresponding to a single LLM call. \\
    $\phi(\mathcal{I}) \in \mathbb{R}^d$ & $d$-dimensional feature embedding of instance $\mathcal{I}$ used for exemplar retrieval. \\
    \addlinespace[4pt]
\end{longtable}

\clearpage
\section{Asset Layout Planning and Synthesis (ALPS) Benchmark}
\label{app:cando-dataset}

\subsection{Data Provenance and Annotation Process}
\label{app:data-provenance}
ALPS-Bench is derived from real operational assets supplied by an industrial setting and was not collected online. An external annotation company assigned approximately $20$ professional annotators under a dedicated project manager. Annotation took place over five weeks, with weekly review sessions in which two domain experts from the asset-design team inspected and refined the annotations over multiple rounds. Only anonymized site geometries are released.

\subsection{Annotation Format}
\label{app:annotation-format}
Annotations are stored in JSON Lines format, one record per layout, each containing a unique identifier and a list of element records. Every element specifies a class label and a polygon attribute as an ordered list of $\mathbb{R}^2$ coordinate pairs in meters, consistent with the formulation in Eq.~\eqref{eq:placed}. The site premise is stored as an element of its own class.

\subsection{Dataset Statistics}
\label{app:data-statistics}
We compute descriptive statistics over all $1{,}000$ canonical layouts in the development and test sets. Every canonical layout has a valid boundary polygon and a paired constraint record. Site premise areas $|\mathcal{P}_i|$ range from $390.33$ to $19{,}358.82$~m$^2$ (mean $3{,}228.40$~m$^2$; median $2{,}705.12$~m$^2$). Excluding the premise record, layouts contain $7$--$101$ polygon annotations (mean $22.75$; median $21$).

\begin{figure}[ht]
  \centering
  \includegraphics[height=0.27\textwidth]{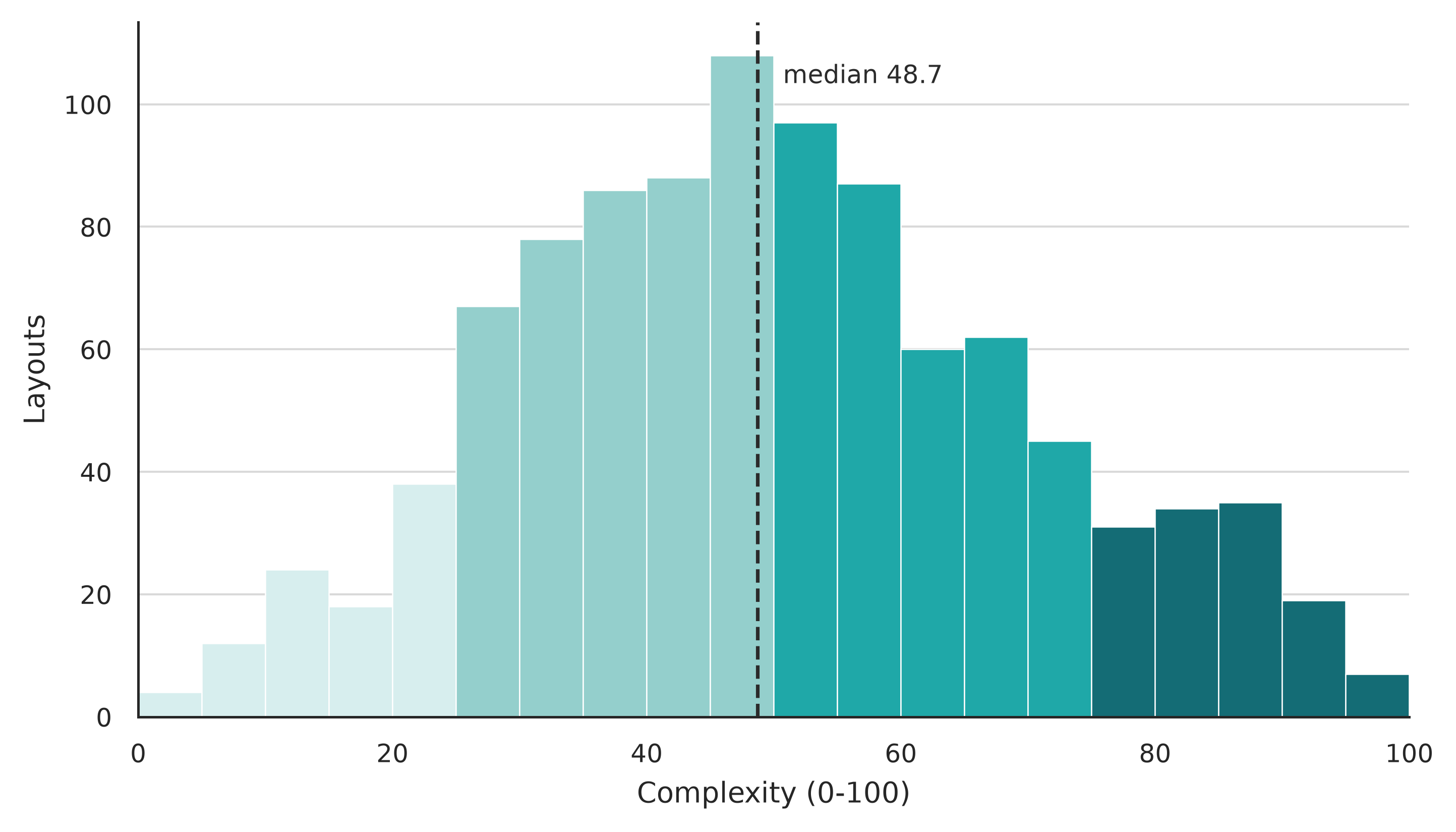}\hfill
  \includegraphics[height=0.27\textwidth]{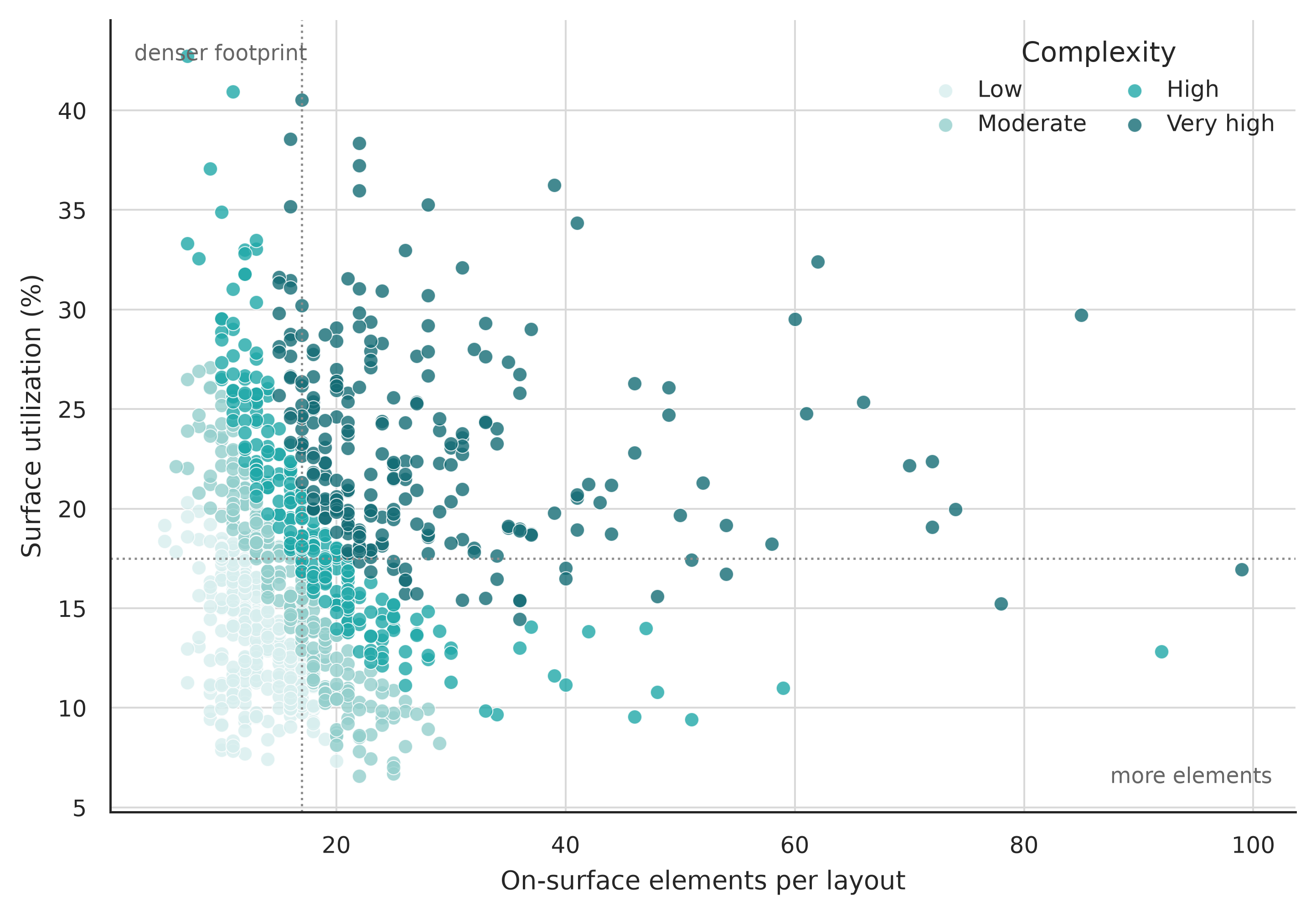}
  \caption{Complexity statistics for \textbf{ALPS-Bench}. \textbf{Left}: Distribution of the surface complexity score. \textbf{Right}: Its two components, surface-annotation count and surface utilization; dotted lines mark their medians. Darker shades indicate greater complexity in both panels.}
  \label{fig:dataset-complexity}
\end{figure}

\paragraph{Surface complexity.}
We characterize surface complexity using two properties: the number of polygon annotations on the surface and the fraction of the premise area they cover. Access regions are included, while the premise boundary and \textsc{Underground Service Facility} are excluded because they do not represent occupied surface facilities. We rank layouts separately by annotation count and covered area, then combine the two ranks with equal weight into a score from $0$ to $100$. This allows annotation count and spatial coverage to contribute equally despite using different units. The median surface utilization is $17.5\%$, and the median complexity score is $48.7$. Figure~\ref{fig:dataset-complexity} shows that layouts may be complex because they contain many surface annotations, cover a large portion of the premise, or both.

\begin{figure}[ht]
  \centering
  \includegraphics[width=\textwidth]{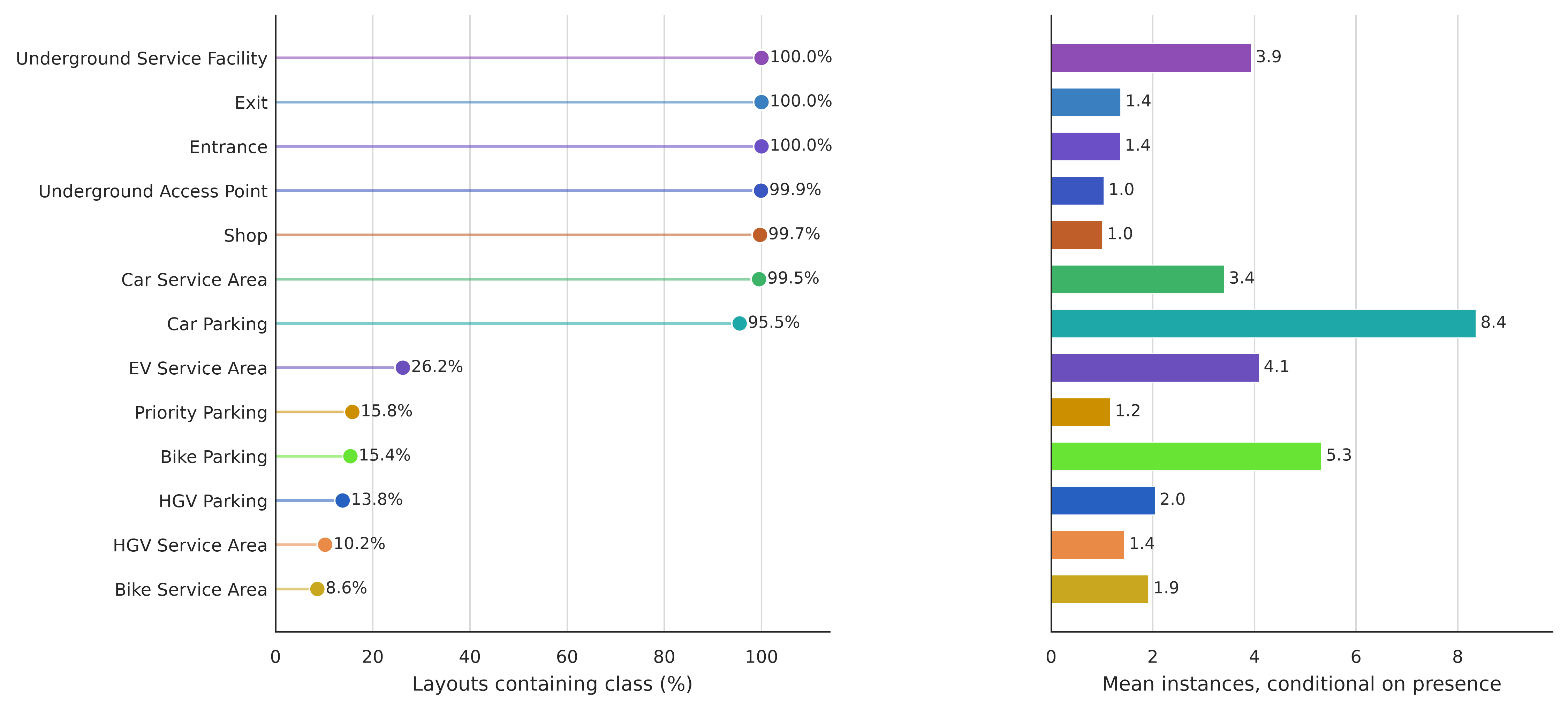}
  \caption{Class distribution in \textbf{ALPS-Bench}. \textbf{Left}: Fraction of layouts containing each class. \textbf{Right}: Mean number of instances among layouts in which the class is present. The premise record is omitted.}
  \label{fig:dataset-class-distribution}
\end{figure}

\paragraph{Class distribution.}
The benchmark exhibits a common operational core: \textsc{Underground Service Facility}, \textsc{Entrance}, and \textsc{Exit} occur in every layout, followed by \textsc{Underground Access Point} ($99.9\%$), \textsc{Shop} ($99.7\%$), \textsc{Car Service Area} ($99.5\%$), and \textsc{Car Parking} ($95.5\%$). Specialized infrastructure is intentionally less frequent: \textsc{EV Service Area} occurs in $26.2\%$ of layouts, \textsc{Priority Parking} in $15.8\%$, \textsc{Bike Parking} in $15.4\%$, \textsc{HGV Parking} in $13.8\%$, \textsc{HGV Service Area} in $10.2\%$, and \textsc{Bike Service Area} in $8.6\%$. Figure~\ref{fig:dataset-class-distribution} reports both class coverage and mean within-layout multiplicity conditional on presence, separating optional capabilities from frequently replicated facilities such as parking.

\subsection{Design Manual}
\label{app:design-manual}
As discussed in Section~\ref{sec:formulation} the design manual $\mathcal{M}$ covers two sets of desired specifications, the \emph{Hard ($\mathcal{HC}$)} and \emph{Soft ($\mathcal{SC}$}) constraints. Let us delve deeper into the underlying rules.

\begin{table}[t]
  \centering
  \caption{Evaluated routing sequences per vehicle type. All routes implicitly begin at any \textsc{Entrance} and terminate at any \textsc{Exit}.}
  \label{tab:routing_scenarios}
  {
  \setlength{\aboverulesep}{0pt}%
  \setlength{\belowrulesep}{0pt}%
  \setlength{\tabcolsep}{10pt}%
  \renewcommand{\arraystretch}{1.25}%
  \small
  \begin{tabular}{@{} ll @{}}
    \specialrule{\heavyrulewidth}{0pt}{0pt}
    \textbf{Vehicle type} & \textbf{Intermediate Waypoint Sequence} \\
    \specialrule{\lightrulewidth}{0pt}{0pt}

    \textbf{Car}
      & \textsc{Car Service Area} \\
      & \textsc{Car Service Area} $\rightarrow$ \textsc{Car Parking} \\
      & \textsc{Car Service Area} $\rightarrow$ \textsc{Priority Parking} \\
      & \textsc{Car Parking} \\
      & \textsc{EV Service Area} \\
    \midrule

    \textbf{Bike}
      & \textsc{Bike Service Area} \\
      & \textsc{Bike Service Area} $\rightarrow$ \textsc{Bike Parking} \\
      & \textsc{Bike Parking} \\
    \midrule

    \textbf{Heavy Goods Vehicle (HGV)}
      & \textsc{HGV Service Area} \\
      & \textsc{HGV Service Area} $\rightarrow$ \textsc{HGV Parking} \\
      & \textsc{HGV Parking} \\

    \bottomrule
  \end{tabular}
  }
\end{table}

\subsubsection{Hard Constraints}
\label{app:hard-constraints}

As established in the problem formulation (Section~\ref{sec:formulation}) and dataset characteristics (Section~\ref{sec:dataset}), valid layouts must comprehensively satisfy fundamental geometric rules. All placed elements must be fully contained within the layout premise $\mathcal{P}$ (containment), and no two placed elements may overlap in their occupied area (non-overlap). Furthermore, the quantities of elements generated must strictly match the problem instance specifications.

Regarding routing feasibility ($\mathrm{HC}_f$), we enforce strict flow requirements to guarantee collision-free maneuverability across the premises. We outline the specifics of these requirements in the preceding and subsequent sections (see Appendix~\ref{app:routing} and Appendix~\ref{app:voronoi}).

\subsubsection{Soft Constraints}
\label{app:soft-constraints}

We distill the operational and aesthetic guidelines of the \textbf{ALPS-Bench} domain into a structured set of soft constraints ($\mathcal{SC}$) that evaluate the structural organization and practical usability of the generated layouts. These are mathematically evaluated as continuous sub-objectives across three central pillars: Compositionality, Proximity, and Orientation. Table~\ref{tab:soft_rules_combined} offers a breakdown summarizing the parameters governing the soft constraints.

\paragraph{Compositionality.}
The Compositionality dimension ensures that multiple instances of identical semantic classes assemble into contiguous, coherent functional zones rather than fragmenting randomly across the layout premise. This is enforced through three foundational parameters: spatial clustering distance thresholds, average group size expectations, and structural arrangement patterns.
The \textbf{Average Group Size} acts as a normalization factor to penalize overly fragmented layouts. For example, if the required average group size for \textsc{Car Service Area} is $4$, distributing $8$ units into one group of $8$ arrays, or two hubs of $4$, yields high compliance, whereas scattering them across $8$ separate $1$-unit islands triggers heavy structural penalties.
Furthermore, clustered elements must configure themselves according to pre-defined geometric patterns. For classes designated as adhering to \textbf{Rectangular or Single Line} topologies, the expected formation dynamic is dictated by the total element count parity within the cluster. Specifically, if the cluster contains an \textit{odd} number of elements (e.g., $5$ units), they must conform strictly to a continuous $1\times5$ or $5\times1$ linear vector grid. Conversely, if the group consists of an \textit{even} number of elements, the system robustly prefers grid-like \textit{rectangular blocks} to maximize space utilization compactness (e.g., configuring $4$ units into a dense $2\times2$ matrix instead of a $1\times4$ strip, or $6$ instances into a $2\times3$ or $3\times2$ block).

\paragraph{Proximity and Clearances.} These dictate topological positioning (e.g., \textsc{shop} structures and \textsc{bike/hgv parking} favoring placement immediately proximally adjacent to, or geometrically stacked against, the outer \textsc{layout polygon} boundary), direct functional bounding (e.g., \textsc{car parking} consistently stacking its shorter geometric edge against the perimeter of a \textsc{shop}), and global distance isolation to mitigate interference (e.g., isolating noisy \textsc{hgv service areas} away from general \textsc{car/bike service areas} using a hard $6.0$m buffer). Most critically, to guarantee navigational flow, elements such as active parking bays require \textbf{Clearance Zones}; stacked structures must maintain an unoccluded outward minimum $180^\circ$ radius scaling to a $3.0$m margin directly abutting the unstacked exit edge, preventing vehicles from being trapped after docking.

\paragraph{Orientation.}
Elements aggregated within a unified spatial group are strictly evaluated for their internal angular agreement. Once a baseline rotational vector is established for a composite functional zone, all adjacent sister elements contributing to that zone must independently match it, either by situating strictly parallel ($0^\circ, 180^\circ$) or perpendicular ($90^\circ, 270^\circ$) to one another axis, capped by a rigid error tolerance threshold of precisely $5^\circ$.

\begin{table}[ht]
  \centering
  \caption{Comprehensive breakdown of $\mathcal{SC}$}
  \label{tab:soft_rules_combined}
  \small
  \begin{tabular}{@{}p{0.18\linewidth}p{0.78\linewidth}@{}}
    \toprule
    \textbf{Constraint Type} & \textbf{Rule Specification} \\
    \midrule
    \multicolumn{2}{@{}l}{\cellcolor{gray!15}\textbf{\textit{Compositionality}}} \\
    \midrule
    Rectangular Blocks \newline/ Single Line & \textsc{car service area} (group size 4, distance tolerance up to 3.0 m), \textsc{bike service area} (group size 2, distance tolerance up to 3.0 m), and \textsc{underground service facility} (group size 2, distance tolerance up to 1.0 m). Odd counts format to primary lines; even counts format to dense multi-axis grids (for example, 2 by 2, or 2 by 3 blocks). \\
    \addlinespace[2pt]
    Strict Single Line & \textsc{car parking} (group size 3, tolerance up to 0.5 m), \textsc{bike parking} (size 4, tolerance up to 0.5 m), \textsc{hgv service area} (size 2, tolerance up to 10.0 m), \textsc{priority parking} (size 2, tolerance up to 1.0 m), \textsc{ev service area} (any size). \\
    \midrule
    \multicolumn{2}{@{}l}{\cellcolor{gray!15}\textbf{\textit{Proximity and Clearances}}} \\
    \midrule
    Boundary Stacking & \textsc{underground access point} (longer edge), \textsc{bike parking}, \textsc{hgv parking}, and \textsc{shop} (longer edge) are ideally stacked against or placed proximally adjacent to the \textsc{layout polygon}. \\
    \addlinespace[2pt]
    Element Stacking & \textsc{car parking} (shorter edge) and \textsc{priority parking} must uniformly stack against the \textsc{shop}. \textsc{priority parking} must also remain placed closest to the \textsc{entrance}. \\
    \addlinespace[2pt]
    Egress Clearances & All parked vehicles (\textsc{car parking}, \textsc{bike parking}, \textsc{hgv parking}, \textsc{priority parking}, and \textsc{ev service area}) necessitate a strictly maintained 3.0 m outward 180 degree maneuvering clearance from their unstacked active exit edge. \\
    \addlinespace[2pt]
    Adjacency \&\newline Distances & \textsc{hgv service area} enforces isolated minimum distances of 6.0 m away from \textsc{car service area} and \textsc{bike service area}. Every distinct service area hub strictly requires at least one \textsc{underground service facility} to be proximally adjacent within 5.0 m to 10.0 m. \\
    \midrule
    \multicolumn{2}{@{}l}{\cellcolor{gray!15}\textbf{\textit{Orientation}}} \\
    \midrule
    Angular Alignment & Elements constituting a unified spatial group must exhibit strict orientational coherence, aligning parallel or perpendicular to one another within a narrow angular tolerance margin of 5 degrees. \\
    \bottomrule
  \end{tabular}
\end{table}

\subsection{Routing Scenarios}
\label{app:routing}
\begin{wraptable}{r}{0.38\textwidth}
  \vspace{-15pt}
  \centering
  \caption{Average vehicle dimensions.}
  \label{tab:vehicle_sizes}
  {%
  \setlength{\tabcolsep}{4pt}
  \renewcommand{\arraystretch}{1.1}
  \small
  \begin{tabular}{@{}lcc@{}}
    \toprule
    \textbf{Vehicle} & \textbf{Length (m)} & \textbf{Width (m)} \\
    \midrule
    Car & 4.5 & 1.8 \\
    Bike & 2.0 & 0.75 \\
    HGV & 16.5 & 2.55 \\
    \bottomrule
  \end{tabular}%
  }
  \vspace{-10pt}
\end{wraptable}

To thoroughly evaluate the operational accessibility and flow of the generated layouts, we define a comprehensive set of routing scenarios across three distinct vehicle types: Cars, Bikes, and Heavy Goods Vehicles (HGVs). Each vehicle type requires a specific minimum path clearance corresponding to its physical dimensions (detailed in Table~\ref{tab:vehicle_sizes}) to ensure collision-free navigation.

A valid routing scenario requires the vehicle to start at any element classified as an \textsc{entrance}, navigate sequentially through a defined list of intermediate facility waypoints, and successfully reach any \textsc{exit} element. Table~\ref{tab:routing_scenarios} details the exhaustive list of specific waypoint sequences evaluated for each vehicle type.

\subsection{Voronoi-Based Motion Planning Evaluation}
\label{app:voronoi}
\begin{figure}[t]
  \centering
  \includegraphics[width=\linewidth]{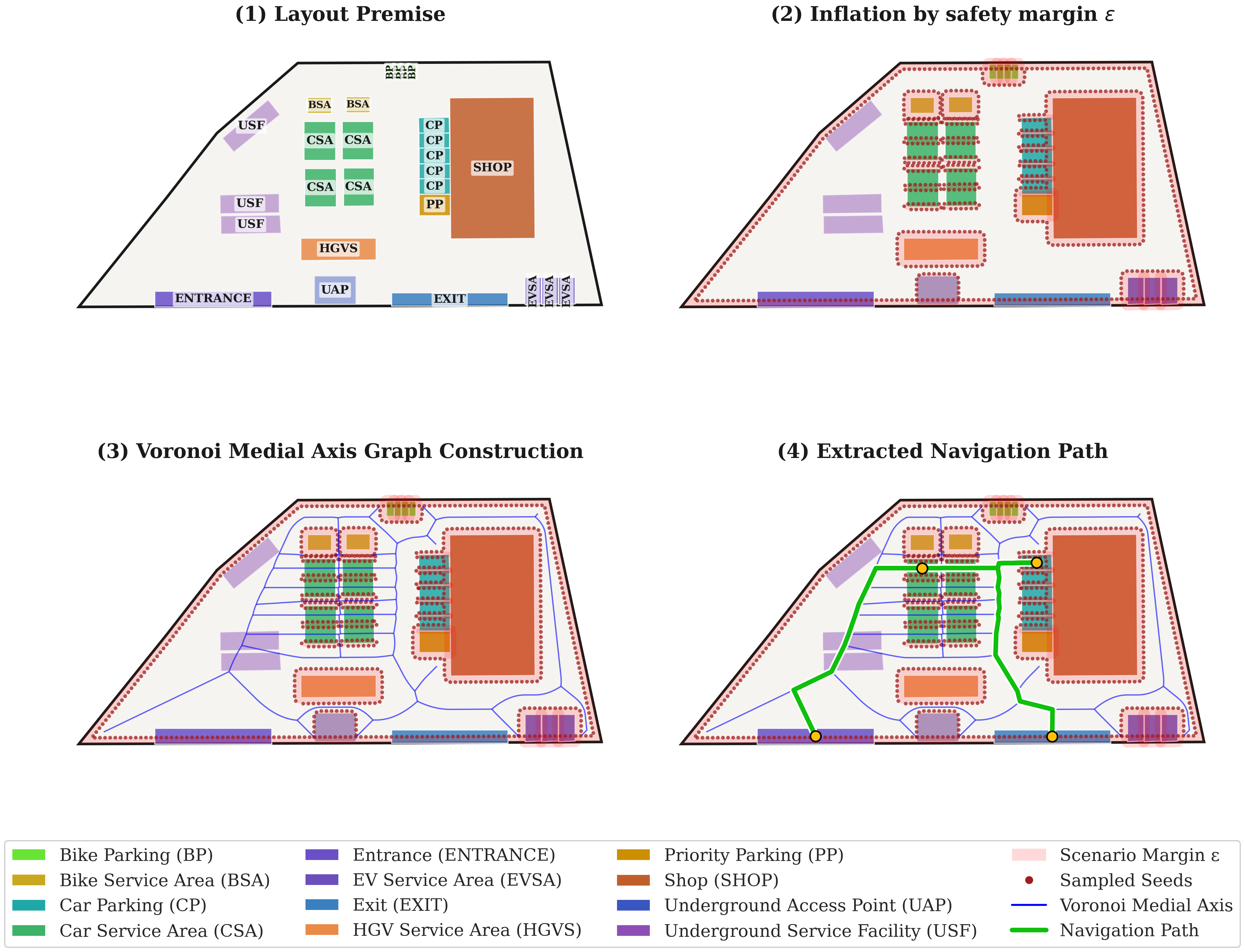}
  \caption{Step-by-step illustration of \texttt{CANDO}'s motion planning tool.}
  \label{app:fig:voronoi_motion_planning}
\end{figure}

Our approach breaks the complex task of navigating a layout into four systematic steps, as illustrated in Figure~\ref{app:fig:voronoi_motion_planning}. First, we establish the initial layout premise for a given scenario (\textbf{Step 1}). Next, we define the free space by inflating all elements that are not part of the scenario (and thus considered \emph{obstacles}) by a safety margin $\epsilon$, with the value dependent on the vehicle size from Table~\ref{tab:vehicle_sizes}. To discretize this continuous margin and map out the available driving area, we sample seed points uniformly across the boundaries of the obstacles (\textbf{Step 2}). This ensures any generated path naturally avoids collisions. We then compute a Voronoi diagram over these points, extracting a medial axis graph (\textbf{Step 3}). This effectively creates a path of maximum clearance from nearby obstacles that runs exactly down the middle of the available corridors. Finally, we map the start and destination coordinates onto this graph and run a shortest-path search to find a smooth and reliable route through the site (\textbf{Step 4}).

Additionally, the motion planning evaluation enforces two specific domain-aware constraints:
\begin{itemize}[leftmargin=*,nosep]
    \item \textbf{Underground Facilities}: Elements classified as \textsc{underground service facility} are assumed to be located below the surface level. Consequently, they are excluded from the scenario-specific obstacle set $\mathcal{O}_r(\mathbf{q})$ and do not obstruct the free space $\mathcal{F}_r(\mathbf{q})$.
    \item \textbf{No-Backtracking Constraints}: For critical operational nodes (\textsc{car service area}, \textsc{bike service area}, and \textsc{hgv service area}), vehicles are strictly prohibited from reversing out the way they entered. They must follow a continuous drive-through trajectory. The only exception occurs if the service area is fully stacked against a site boundary segment, in which case the natural drive-through is constrained by the perimeter.
\end{itemize}

\clearpage
\section{Experiments}
\label{app:experiments}

\subsection{Trained ALPS-Bench Baselines}
\label{app:trained-baselines}

\paragraph{LayoutDM.}
We trained LayoutDM~\cite{inoue2023layoutdm} from scratch on the ALPS-Bench development set using one H100 GPU, 32~GB of host RAM, and a runtime budget of up to 48 hours. We retained its constrained discrete-diffusion formulation, using 100 diffusion steps and $k$-means quantization. Because LayoutDM natively represents axis-aligned page elements as $(x,y,w,h)$, we converted each element polygon to a normalized oriented bounding box in the coordinate frame of its site boundary and extended the token sequence to $(c,w,h,x,y,o)$, where $o$ denotes element orientation. The site boundary and \textsc{Entrance}/\textsc{Exit} elements are placed first in the sequence as fixed context and masked from the prediction loss; orientation loss is also masked for geometrically ambiguous, near-square elements. This adaptation generalizes the tokenizer, diffusion transitions, coordinate clustering, and data pipeline to the orientation variable while retaining the original FID pathway over $(x,y,w,h)$. Consequently, FID remains comparable to the original implementation but does not evaluate orientation. As the oriented-box representation only approximates the true polygonal geometry, generated placements may still cross the site boundary or overlap, explaining LayoutDM's low geometric constraint satisfaction.

\paragraph{Qwen-LoRA-ALPS.}
We fine-tuned Qwen3-VL-8B-Instruct~\cite{yang2025qwen3technicalreport} on the ALPS-Bench development set using parameter-efficient LoRA SFT~\cite{hu2021lora} on one H100 GPU. The base vision encoder, multi-modal projector, and language-model weights remained frozen, while LoRA adapters with rank 8, scaling factor 16, and dropout 0.05 were trained for 30 epochs with per-device batch size 1, a cosine-decayed learning rate of $5\!\times\!10^{-5}$, BF16/TF32 precision, and gradient checkpointing. The maximum input sequence length was 8,192 tokens. Each input combines a rendering of the site boundary and access points with structured element requirements, and the target contains all element polygons in a single response. This setup captures high-level composition and orientation patterns but provides no iterative mechanism for repairing containment, non-overlap, or routing violations.

\subsection{Simplified \texttt{CANDO} for Public Benchmarks}
\label{app:simplified-cando}

All public-benchmark experiments use Qwen3.6-35B-A3B-FP8 (thinking disabled) for every agent role without task-specific fine-tuning. The model is served through OpenAI-compatible vLLM replicas, each hosted on one H100-80GB GPU. The task-aware ICL Initializer retrieves $8$ training-set exemplars and samples $10$ initial layouts, retrying an invalid batch at most three times. Valid layouts are ranked using weights $0.2$, $0.2$, and $0.6$ for alignment, overlap, and dissimilarity from development layouts, respectively.

The simplified pipeline retains the Verifier, Planner, Generator, and Historian, but replaces ALPS-specific routing and compositionality workers with benchmark-specific checks. At each refinement step, deterministic workers evaluate hard constraints, alignment/assembly, and proximity; the Generator samples $K{=}8$ candidate action sequences from seven diverse repair strategies. PubLayNet and RICO use $N{=}4$ parallel branches, $B{=}5$ steps per branch, a maximum of $T{=}10$ iterations, local patience of $3$, global plateau patience of $6$, and a history window of $W{=}9$. Their Verifier categories are weighted uniformly, while Planner guidance weights hard constraints, assembly, and proximity by $0.4$, $0.4$, and $0.2$. PubLayNet additionally uses an LLM-based element-role and single-/double-column classifier, whereas RICO uses its element classes directly.

PKU-PosterLayout uses the same settings except for $N{=}2$ branches. It adds multi-modal content-Verifier and content-Planner agents that inspect the generated layout overlaid on the in-painted poster and saliency map. The original composed poster is excluded to prevent target-layout leakage. Hard constraints, content, assembly, and proximity receive Verifier weights $0.25$, $0.50$, $0.25$, and $0.00$, respectively, and compliance-based termination requires a content score of at least $0.8$. Its strategy library replaces the proximity-only strategy with a content-only strategy.

\begin{table}[ht]
\centering
\caption{Refinement configuration of simplified \texttt{CANDO} on public benchmarks.}
\label{tab:simplified-cando}
    \resizebox{0.78\linewidth}{!}{%
        \begin{tabular}{lccc}
            \toprule
            \textbf{Hyperparameter} & \textbf{PubLayNet} & \textbf{RICO} & \textbf{PKU-PosterLayout} \\
            \midrule
            Retrieved exemplars & $8$ & $8$ & $8$ \\
            Initial layouts & $10$ & $10$ & $10$ \\
            Branches $N$ & $4$ & $4$ & $2$ \\
            Steps per branch $B$ & $5$ & $5$ & $5$ \\
            Candidates per step $K$ & $8$ & $8$ & $8$ \\
            Maximum iterations $T$ & $10$ & $10$ & $10$ \\
            Local / global patience & $3$ / $6$ & $3$ / $6$ & $3$ / $6$ \\
            History window $W$ & $9$ & $9$ & $9$ \\
            \bottomrule
        \end{tabular}%
    }
\end{table}

\paragraph{Evaluation protocol.}
PubLayNet and RICO are evaluated once per sample on their complete test splits, comprising $10{,}998$ and $3{,}729$ trials per task, respectively. We report maximum layout similarity (mIoU), alignment, overlap, and LayoutNet FID with dataset-specific checkpoints. Following prior RICO implementations, overlap excludes the \textsc{Card}, \textsc{Background}, and \textsc{Modal} classes because they function as containers or backgrounds. PKU-PosterLayout uses $1,000$ test samples per task and the RALF protocol: occlusion, unreadability, strict underlay effectiveness, non-underlay overlay, and FIDNetV3 FID. Following the official protocol, elements with normalized area at most $0.001$ are discarded before evaluation, and FID uses the official \texttt{pku10} checkpoint.

\subsection{Compute Resources and Latency}
\label{app:compute}
The ALPS-Bench Qwen experiments used eight NVIDIA H100 GPUs (80\,GB VRAM each), with one Qwen3.6-35B-A3B-FP8 vLLM replica per GPU and 24 concurrent evaluation workers. The GPT-5.2 experiments used the Azure OpenAI API and no local GPUs. The Single-Agent and \texttt{CANDO} CoDe evaluations used 100 concurrent workers, whereas \texttt{CANDO} BoN used 50. LayoutPrompter used the same API infrastructure as the other GPT-5.2 experiments.

Results reported in Table~\ref{tab:compute_resources} distinguish \emph{per-trial latency}, the elapsed time from the start to completion of an individual layout under concurrent load, from \emph{batch wall time}, the makespan of the parallel evaluation over the 100 layouts in $\mathcal{D}^{\textup{test}}$. The \emph{effective time per layout} corresponds to amortized throughput. For local experiments, GPU-hours are computed as the batch wall time multiplied by the number of allocated GPUs. Per-trial durations must therefore not be summed when trials overlap. All latency measurements were collected under concurrent load rather than as isolated single-request measurements.

\begin{table}[ht]
\centering
\caption{Compute resources and latency for final ALPS-Bench test-set evaluations. Mean trial latency is reported as mean $\pm$ standard deviation. Effective time per layout measures throughput, whereas trial latency measures the end-to-end duration of an individual trial under concurrent load.}
\label{tab:compute_resources}
    \resizebox{\linewidth}{!}{%
        \begin{tabular}{llrrrr}
            \toprule
            \textbf{Method} & \textbf{Compute} & \textbf{Mean trial latency} & \textbf{Median} & \textbf{Batch wall time} & \textbf{Effective time/layout} \\
            \midrule
            Single-Agent (GPT-5.2) & Cloud API & $10.49 \pm 5.83$\,h & $12.42$\,h & $20.20$\,h & $\sim$12\,min\,07\,s \\
            \texttt{CANDO} (BoN/GPT-5.2) & Cloud API & $47.12 \pm 39.34$\,min & $33.51$\,min & $2.50$\,h & $\sim$1\,min\,30\,s \\
            \texttt{CANDO} (CoDe/GPT-5.2) & Cloud API & $42.75 \pm 36.34$\,min & $29.42$\,min & $2.39$\,h & $\sim$1\,min\,26\,s \\
            \texttt{CANDO} (CoDe/Qwen) & 8$\times$H100 (80\,GB) & $2.86 \pm 1.73$\,h & $2.67$\,h & $16.7$\,h & $\sim$10\,min\,02\,s \\
            \bottomrule
        \end{tabular}%
    }
\end{table}

Note that the totals in Table~\ref{tab:compute_resources} cover only the final 100-layout ALPS-Bench test evaluations. The broader research effort also included component and hyperparameter ablation trials, reported in Tables~\ref{tab:component_ablation} and~\ref{tab:hyperparameter_tuning}, which we estimate consumed approximately $2,500$ allocated H100 GPU-hours.

\begin{table}[ht]
\centering
\caption{Median per-sample response time of \texttt{CANDO} on the PubLayNet, RICO, and PKU-PosterLayout public benchmarks. Results were measured in isolated runs (single-sample inference, diagnostics disabled) and do not include graph loading overhead.}
\label{tab:public_latency}
    \begin{tabular}{lrrr}
        \toprule
        \textbf{Task} & \textbf{PubLayNet} & \textbf{RICO} & \textbf{PKU} \\
        \midrule
        Gen-T      & 104.8\,s &  66.8\,s &  46.8\,s \\
        Gen-TS     & 125.7\,s & 137.4\,s & 111.4\,s \\
        Gen-R      & 166.1\,s &  74.7\,s & 102.9\,s \\
        Completion &  65.4\,s &  31.1\,s &  18.0\,s \\
        Refinement &  71.8\,s &  54.1\,s & 39.3\,s \\
        \bottomrule
    \end{tabular}
\end{table}

For PubLayNet, RICO, and PKU-PosterLayout, end-to-end latency is measured per sample from initial generation through all candidate evaluations, Verifier calls, and refinement steps. Measurements were obtained in isolated benchmarking runs processing one sample at a time (from a randomly selected subset of $50$ samples per task), with diagnostics disabled. Reported latencies exclude graph loading overhead. Table~\ref{tab:public_latency} reports median latency. We cannot provide matched latency for the competing methods because their reported results were not reproduced in our environment and their original papers do not report elapsed time.

\subsection{Additional Results on PubLayNet and RICO}
\label{app:pubrico-additional-results}

Table~\ref{app:results_publaynet_rico} extends the evaluation in Section~\ref{sec:experiments}, reporting performance on the Layout Relationship Generation (Gen-R), Layout Completion (Comp.), and Layout Refinement (Refine) tasks for both the PubLayNet and RICO benchmarks.

\begin{table}[ht]
\centering
\caption{Quantitative results on the PubLayNet and RICO datasets for Type Generation, Type-Size Generation, Relationship Generation (Gen-R), Completion (Comp.), and Refinement (Refine) tasks.}
\label{app:results_publaynet_rico}
\resizebox{\linewidth}{!}{
    \begin{tabular}{ll cccc c cccc}
        \toprule
        & & \multicolumn{4}{c}{\cellcolor{ricoColor}\textbf{RICO}} && \multicolumn{4}{c}{\cellcolor{publayColor}\textbf{PubLayNet}} \\
        \cmidrule{3-6} \cmidrule{8-11}
        \textbf{Task} & \textbf{Method} & \textbf{mIoU}$\uparrow$ & \textbf{FID}$\downarrow$ & \textbf{Align}$\downarrow$ & \textbf{Over}$\downarrow$ && \textbf{mIoU}$\uparrow$ & \textbf{FID}$\downarrow$ & \textbf{Align}$\downarrow$ & \textbf{Over}$\downarrow$ \\
        \midrule
        \multirow{6}{*}{\textbf{Gen-T}}
        & LayoutGAN++~\citep{layoutgan++} & 0.298 & 5.954 & 0.261 & 0.620 && 0.297 & 14.875 & 0.124 & 0.148 \\
        & BLT~\citep{kong2022blt} & 0.216 & 25.633 & 0.150 & 0.983 && 0.140 & 38.684 & 0.036 & 0.196 \\
        & LayoutFormer++~\citep{jiang2023layoutformer++} & 0.432 & \textbf{1.096} & 0.230 & 0.530 && 0.348 & 8.411 & 0.020 & 0.008 \\
        & LayoutPrompter~\citep{lin2023layoutprompter} & 0.604 & 2.134 & 0.002 & 0.029 && 0.619 & \underline{1.648} & 0.001 & 0.004 \\
        & LayoutCoT~\citep{shi2025layoutcot} & \underline{0.719} & 2.189 & \underline{0.001} & \underline{0.022} && \underline{0.772} & \textbf{1.227} & \textbf{0.000} & \underline{0.002} \\
        \cmidrule{2-11}
        & \cellcolor{candoBlue!10}\texttt{CANDO} (ours) & \cellcolor{candoBlue!10}\textbf{0.866} & \cellcolor{candoBlue!10}\underline{2.101} & \cellcolor{candoBlue!10}\textbf{0.000} & \cellcolor{candoBlue!10}\textbf{0.005} && \cellcolor{candoBlue!10}\textbf{0.893} & \cellcolor{candoBlue!10}3.455 & \cellcolor{candoBlue!10}\textbf{0.000} & \cellcolor{candoBlue!10}\textbf{0.000} \\
        \midrule
        \multirow{5}{*}{\textbf{Gen-TS}}
        & BLT~\citep{kong2022blt} & 0.604 & \underline{0.951} & 0.181 & 0.660 && 0.428 & 7.914 & 0.021 & 0.419 \\
        & LayoutFormer++~\citep{jiang2023layoutformer++} & 0.620 & \textbf{0.757} & 0.202 & 0.542 && 0.471 & \textbf{0.720} & 0.024 & 0.037 \\
        & LayoutPrompter~\citep{lin2023layoutprompter} & 0.707 & 3.072 & 0.002 & \underline{0.051} && 0.727 & 1.863 & 0.001 & 0.076 \\
        & LayoutCoT~\citep{shi2025layoutcot} & \underline{0.795} & 3.506 & \textbf{0.000} & \textbf{0.031} && \underline{0.847} & \underline{1.430} & \textbf{0.000} & \underline{0.004} \\
        \cmidrule{2-11}
        & \cellcolor{candoBlue!10}\texttt{CANDO} (ours) & \cellcolor{candoBlue!10}\textbf{0.879} & \cellcolor{candoBlue!10}1.371 & \cellcolor{candoBlue!10}\textbf{0.000} & \cellcolor{candoBlue!10}0.052 && \cellcolor{candoBlue!10}\textbf{0.881} & \cellcolor{candoBlue!10}2.023 & \cellcolor{candoBlue!10}\textbf{0.000} & \cellcolor{candoBlue!10}\textbf{0.000} \\
        \midrule
        \multirow{4}{*}{\textbf{Gen-R}}
        & LayoutFormer++~\cite{jiang2023layoutformer++} & 0.424 & 5.972 & 0.332 & 0.537 && 0.353 & 4.954 & 0.025 & 0.076 \\
        & LayoutPrompter~\cite{lin2023layoutprompter} & 0.583 & \textbf{2.249} & \underline{0.001} & 0.044 && 0.627 & \underline{2.224} & \underline{0.001} & 0.011 \\
        & LayoutCoT~\cite{shi2025layoutcot} & \underline{0.725} & 3.569 & \underline{0.001} & \underline{0.015} && \underline{0.769} & \textbf{1.408} & \underline{0.001} & \textbf{0.002} \\
        \cmidrule{2-11}
        & \cellcolor{candoBlue!10}\texttt{CANDO} (ours) & \cellcolor{candoBlue!10}\textbf{0.844} & \cellcolor{candoBlue!10}\underline{2.674} & \cellcolor{candoBlue!10}\textbf{0.000} & \cellcolor{candoBlue!10}\textbf{0.009} && \cellcolor{candoBlue!10}\textbf{0.855} & \cellcolor{candoBlue!10}9.614 & \cellcolor{candoBlue!10}\textbf{0.000} & \cellcolor{candoBlue!10}\textbf{0.002} \\
        \midrule
        \multirow{4}{*}{\textbf{Comp.}}
        & LayoutFormer++~\cite{jiang2023layoutformer++} & \underline{0.732} & 4.574 & 0.077 & 0.487 && 0.471 & 10.25 & 0.020 & 0.002 \\
        & LayoutPrompter~\cite{lin2023layoutprompter} & 0.541 & 2.656 & \underline{0.001} & 0.041 && 0.754 & \underline{1.282} & \underline{0.001} & 0.001 \\
        & LayoutCoT~\cite{shi2025layoutcot} & 0.716 & \underline{2.112} & \underline{0.001} & \underline{0.012} && \underline{0.770} & \textbf{0.531} & \textbf{0.000} & \underline{0.001} \\
        \cmidrule{2-11}
        & \cellcolor{candoBlue!10}\texttt{CANDO} (ours) & \cellcolor{candoBlue!10}\textbf{0.894} & \cellcolor{candoBlue!10}\textbf{1.717} & \cellcolor{candoBlue!10}\textbf{0.000} & \cellcolor{candoBlue!10}\textbf{0.002} && \cellcolor{candoBlue!10}\textbf{0.912} & \cellcolor{candoBlue!10}1.904 & \cellcolor{candoBlue!10}\textbf{0.000} & \cellcolor{candoBlue!10}\textbf{0.000} \\
        \midrule
        \multirow{4}{*}{\textbf{Refine}}
        & LayoutFormer++~\cite{jiang2023layoutformer++} & 0.816 & \textbf{0.032} & 0.123 & 0.489 && 0.785 & \textbf{0.086} & 0.024 & 0.006 \\
        & LayoutPrompter~\cite{lin2023layoutprompter} & 0.874 & \underline{0.225} & 0.001 & 0.110 && 0.756 & 0.143 & 0.003 & 0.001 \\
        & LayoutCoT~\cite{shi2025layoutcot} & \underline{0.890} & 0.353 & \textbf{0.000} & \underline{0.052} && \underline{0.839} & \underline{0.141} & \textbf{0.000} & \textbf{0.000} \\
        \cmidrule{2-11}
        & \cellcolor{candoBlue!10}\texttt{CANDO} (ours) & \cellcolor{candoBlue!10}\textbf{0.911} & \cellcolor{candoBlue!10}0.370 & \cellcolor{candoBlue!10}\textbf{0.000} & \cellcolor{candoBlue!10}\textbf{0.008} && \cellcolor{candoBlue!10}\textbf{0.941} & \cellcolor{candoBlue!10}0.279 & \cellcolor{candoBlue!10}\textbf{0.000} & \cellcolor{candoBlue!10}\textbf{0.000} \\
        \bottomrule
    \end{tabular}
}
\end{table}

\subsection{Additional Results on PKU-PosterLayout}
\label{app:pku-additional-results}

Table~\ref{app:results_pku} extends the evaluation in Section~\ref{sec:experiments}, reporting performance on the Type Generation (Gen-T), Type-Size Generation (Gen-TS), Layout Relationship Generation (Gen-R), Layout Completion (Comp.), and Layout Refinement (Refine) tasks for the PKU-PosterLayout benchmark.

\begin{table}[ht]
\centering
\caption{Quantitative results on the PKU-PosterLayout dataset for Type Generation (Gen-T), Type-Size Generation (Gen-TS), Relationship Generation (Gen-R), Completion (Comp.), and Refinement (Refine) tasks.} \label{app:results_pku}
\resizebox{0.7\linewidth}{!}{
    \begin{tabular}{ll ccccc}
        \toprule
        \textbf{Task} & \textbf{Method} & \textbf{Occ}$\downarrow$ & \textbf{Rea}$\downarrow$ & \textbf{Und$_S$}$\uparrow$ & \textbf{Ove}$\downarrow$ & \textbf{FID}$\downarrow$ \\
        \midrule
        \multirow{3}{*}{\textbf{Gen-T}} & RALF & \underline{0.136} & \textbf{0.0127} & 0.890 & 0.0091 & \textbf{2.36} \\ & UniLayDiff & \textbf{0.122} & \underline{0.0129} & \underline{0.963} & \underline{0.0018} & \underline{3.09} \\ \cmidrule{2-7} & \cellcolor{candoBlue!10}\texttt{CANDO} (ours) & \cellcolor{candoBlue!10}0.144 & \cellcolor{candoBlue!10}0.0196 & \cellcolor{candoBlue!10}\textbf{1.000} & \cellcolor{candoBlue!10}\textbf{0.0009} & \cellcolor{candoBlue!10}9.83 \\
        \midrule
        \multirow{3}{*}{\textbf{Gen-TS}} & RALF & \underline{0.138} & \textbf{0.0123} & 0.876 & 0.0113 & \textbf{0.61} \\ &
        UniLayDiff & \textbf{0.128} & \underline{0.0132} & \underline{0.918} & \underline{0.0046} & \underline{0.66} \\
        \cmidrule{2-7} & \cellcolor{candoBlue!10}\texttt{CANDO} (ours) & \cellcolor{candoBlue!10}0.145 & \cellcolor{candoBlue!10}0.0197 & \cellcolor{candoBlue!10}\textbf{0.990} & \cellcolor{candoBlue!10}\textbf{0.0000} & \cellcolor{candoBlue!10}4.47 \\
        \midrule
        \multirow{3}{*}{\textbf{Gen-R}} & RALF & \underline{0.124} & \underline{0.0138} & 0.869 & 0.0085 & \underline{2.41} \\ & UniLayDiff & \textbf{0.119} & \textbf{0.0131} & \underline{0.883} & \underline{0.0049} & \textbf{2.15} \\
        \cmidrule{2-7} &
        \cellcolor{candoBlue!10}\texttt{CANDO} (ours) & \cellcolor{candoBlue!10}0.154 & \cellcolor{candoBlue!10}0.0206 & \cellcolor{candoBlue!10}\textbf{0.990} & \cellcolor{candoBlue!10}\textbf{0.0000} & \cellcolor{candoBlue!10}9.95 \\
        \midrule
        \multirow{3}{*}{\textbf{Comp.}} & RALF & \underline{0.137} & \underline{0.0132} & 0.905 & 0.0139 & \textbf{1.46} \\ & UniLayDiff & \textbf{0.116} & \textbf{0.0128} & \underline{0.974} & \underline{0.0021} & \underline{2.69} \\
        \cmidrule{2-7} &
        \cellcolor{candoBlue!10}\texttt{CANDO} (ours) & \cellcolor{candoBlue!10}\textbf{0.116} & \cellcolor{candoBlue!10}0.0159 & \cellcolor{candoBlue!10}\textbf{1.000} & \cellcolor{candoBlue!10}\textbf{0.0000} & \cellcolor{candoBlue!10}5.25 \\
        \midrule
        \multirow{3}{*}{\textbf{Refine}} & RALF & 0.128 & \underline{0.0103} & \underline{0.951} & 0.0047 & \textbf{0.11} \\ & UniLayDiff & \underline{0.123} & \textbf{0.0102} & \textbf{0.973} & \underline{0.0010} & \underline{0.76} \\
        \cmidrule{2-7} &
        \cellcolor{candoBlue!10}\texttt{CANDO} (ours) & \cellcolor{candoBlue!10}\textbf{0.100} & \cellcolor{candoBlue!10}0.0139 & \cellcolor{candoBlue!10}0.920 & \cellcolor{candoBlue!10}\textbf{0.0000} & \cellcolor{candoBlue!10}3.00 \\
        \bottomrule
    \end{tabular}
}
\end{table}

\subsection{Component Ablation Analysis}
\label{app:component-ablation}

We provide additional analysis of the architectural ablations reported in Table~\ref{tab:component_ablation}. In particular, we examine the effect of the strategy-diversification mechanism used during refinement.

\textbf{Strategy Diversification.} While the average AGG and CS scores remain relatively stable when removing strategy diversification (Table~\ref{tab:component_ablation}), the regression rate increases substantially from $7.0\%$ to $18.4\%$. The regression rate measures the percentage of refinement iterations in which the verification score decreases after applying the proposed changes. This suggests that relying on a single generic refinement instruction makes the system considerably more likely to resolve one violation while inadvertently introducing or worsening another.

\begin{table}[ht]
\centering
\caption{Distribution of winning refinement strategies across \texttt{CANDO} refinement steps. A strategy is considered the winner when its candidate achieves the highest verification score among the strategies explored at that step.}
\label{tab:strategy_distribution}
\resizebox{0.55\linewidth}{!}{
    \begin{tabular}{l cc}
    \toprule
    \textbf{Strategy} & \textbf{\# Times Won} & \textbf{Win Percentage} \\
    \midrule
    Conservative & 930 & 17.6\% \\
    Proximity-focused & 875 & 16.6\% \\
    Surgical & 724 & 13.7\% \\
    Aggressive & 629 & 11.9\% \\
    Macro-spatial & 610 & 11.5\% \\
    Compositionality-focused & 548 & 10.4\% \\
    Space-utilization & 531 & 10.0\% \\
    Hard-constraint-focused & 439 & 8.3\% \\
    \bottomrule
    \end{tabular}
}
\end{table}

Table~\ref{tab:strategy_distribution} further illustrates the benefit of maintaining multiple refinement strategies. No single strategy dominates the search process: the most frequently selected strategy, Conservative, accounts for only $17.6\%$ of winning candidates, followed closely by Proximity-focused at $16.6\%$, while even the least frequently selected Hard-constraint-focused strategy accounts for $8.3\%$. The relatively broad distribution indicates that different constraint configurations require qualitatively different repair behaviors. This supports the design choice of exploring several specialized refinement strategies in parallel rather than relying on a single fixed instruction.

\subsection{Hyperparameter Tuning}
\label{app:hyperparameter-ablation}

Table~\ref{tab:hyperparameter_tuning} analyzes the joint impact of the number of branches $N$, block size $B$, per-step candidates $K$, and history window $W$, using $T{=}10$ during hyperparameter selection to control search cost. The best tested setting is $N{=}4$, $B{=}10$, $K{=}6$, and $W{=}6$, achieving $0.76$ CS, $0.42$ RE, and $0.59$ AGG. We use this setting with a final evaluation budget of $T{=}20$ in the reported ALPS-Bench experiments. Configurations with more branches ($N{=}4$) and larger blocks ($B{\ge}5$) tend to perform best, supporting prolonged multi-path exploration over greedy, single-step commitments. For matched $N$, $B$, and $K$, increasing $W$ from $3$ to $6$ improves AGG from $0.52$ to $0.54$, highlighting the Planner's use of historical trajectories to avoid cyclic edits.

\begin{table}[ht]
\centering
\caption{Impact of $N$, $B$, $K$, and $W$, for $T{=}10$.}
\label{tab:hyperparameter_tuning}
\resizebox{0.55\linewidth}{!}{
  \begin{tabular}{cccc l ccc}
  \toprule
  \multicolumn{4}{c}{\textbf{Hyperparameter}} && \multicolumn{3}{c}{\textbf{Evaluation Metric}} \\
  \cmidrule{1-4} \cmidrule{6-8}
  \textbf{$N$} & \textbf{$B$} & \textbf{$K$} & \textbf{$W$} && \textbf{CS}$\uparrow$ & \textbf{RE}$\uparrow$ & \textbf{AGG}$\uparrow$ \\
  \midrule
  4 & 10 & 6 & 6 && \textbf{0.76} & \textbf{0.42} & \textbf{0.59} \\
  4 & 5  & 8 & 6 && 0.71 & 0.37 & 0.54 \\
  2 & 5  & 4 & 6 && 0.68 & 0.37 & 0.52 \\
  4 & 5  & 8 & 3 && 0.67 & 0.37 & 0.52 \\
  2 & 10 & 8 & 9 && 0.63 & 0.38 & 0.51 \\
  4 & 10 & 4 & 3 && 0.63 & 0.32 & 0.48 \\
  1 & 5  & 8 & 3 && 0.59 & 0.33 & 0.46 \\
  2 & 1  & 6 & 6 && 0.58 & 0.33 & 0.46 \\
  2 & 10 & 4 & 9 && 0.57 & 0.34 & 0.46 \\
  1 & 5  & 4 & 9 && 0.58 & 0.30 & 0.44 \\
  1 & 5  & 8 & 6 && 0.60 & 0.27 & 0.43 \\
  \bottomrule
  \end{tabular}
}
\end{table}

\paragraph{Block Size.}
\begin{table}[t]
\centering
\caption{Effect of block size $B$ on best layout score, constraint compliance rate, and mean number of iterations to comply. Fixed hyperparameters: $N{=}4$, $K{=}8$, $W{=}6$, $T_{\max}{=}20$.}
\label{app:block_size_ablation}
\resizebox{0.6\linewidth}{!}{
    \begin{tabular}{c cc c c}
    \toprule
    \textbf{Block Size} ($B$) & \textbf{CS}$\uparrow$ & \textbf{Compliance Rate}$\uparrow$ & \textbf{Mean Iters to Comply}$\downarrow$ \\
    \midrule
    1  & $0.702_{\pm 0.274}$ & 30.6\% & 7.0 \\
    5  & $\mathbf{0.750}_{\pm 0.264}$ & 38.0\% & 6.9 \\
    8  & $0.730_{\pm 0.271}$ & 40.0\% & 6.1 \\
    10 & $0.742_{\pm 0.277}$ & 36.4\% & 5.8 \\
    20 & $0.740_{\pm 0.270}$ & \textbf{42.4\%} & 8.3 \\
    \bottomrule
    \end{tabular}
}
\end{table}
Table~\ref{app:block_size_ablation} reports the effect of varying the block size $B$, while keeping all other hyperparameters fixed. The relationship between $B$ and layout quality is non-monotonic: performance peaks at $B{=}5$ (mean score $0.750$) and degrades for both smaller and larger values. Small block sizes ($B{=}1$) limit the Generator's ability to reason about multi-step spatial dependencies within a single call, leading to the lowest mean score and compliance rate. At the other extreme, large block sizes ($B{=}20$) yield the highest constraint compliance rate (42.4\%) but at the cost of longer iteration sequences and a slightly lower best score. This suggests a bias--variance trade-off: smaller blocks produce faster but less coherent placements. Because this one-factor analysis fixes $K{=}8$, it does not determine the final joint configuration: $B{=}5$ maximizes mean CS in this slice, whereas the joint sweep in Table~\ref{tab:hyperparameter_tuning} selects $B{=}10$ together with $K{=}6$.

\subsection{Statistical Significance on ALPS-Bench}
\label{app:significance-testing}

We complement the summary statistics in Table~\ref{tab:cando_test_results} with paired, instance-level significance testing over the same $100$ held-out instances. We conduct two-sided Wilcoxon signed-rank tests with Holm correction for multiple comparisons and compute paired-bootstrap $95\%$ confidence intervals (CIs). Table~\ref{tab:significance_alps} reports the mean paired difference $\Delta$ (first method minus second), its $95\%$ CI, and the corresponding Holm-adjusted $p$-value for AGG, CS, and RE.

\begin{table*}[ht]
\centering
\caption{Paired significance tests on \textbf{ALPS-Bench} $\mathcal{D}^{\text{test}}$ ($N{=}100$) using two-sided Wilcoxon signed-rank tests with Holm correction and paired-bootstrap $95\%$ confidence intervals. $\Delta$ denotes the mean paired difference (first method minus second).}
\label{tab:significance_alps}
\resizebox{\textwidth}{!}{
    \begin{tabular}{l ccc ccc ccc}
        \toprule
        & \multicolumn{3}{c}{\textbf{AGG}} & \multicolumn{3}{c}{\textbf{CS}} & \multicolumn{3}{c}{\textbf{RE}} \\
        \cmidrule{2-4} \cmidrule{5-7} \cmidrule{8-10}
        \textbf{Comparison} & $\boldsymbol{\Delta}$ & \textbf{95\% CI} & $\boldsymbol{p_{\mathrm{Holm}}}$ & $\boldsymbol{\Delta}$ & \textbf{95\% CI} & $\boldsymbol{p_{\mathrm{Holm}}}$ & $\boldsymbol{\Delta}$ & \textbf{95\% CI} & $\boldsymbol{p_{\mathrm{Holm}}}$ \\
        \midrule
        CANDO-Qwen vs. Qwen-LoRA-ALPS & 0.543 & [0.497, 0.585] & $<0.001$ & 0.816 & [0.762, 0.864] & $<0.001$ & 0.281 & [0.228, 0.332] & $<0.001$ \\
        CANDO-GPT vs. Qwen-LoRA-ALPS & 0.527 & [0.473, 0.578] & $<0.001$ & 0.768 & [0.697, 0.835] & $<0.001$ & 0.289 & [0.237, 0.341] & $<0.001$ \\
        \midrule
        CANDO-Qwen vs. LayoutDM & 0.491 & [0.451, 0.529] & $<0.001$ & 0.822 & [0.770, 0.870] & $<0.001$ & 0.170 & [0.120, 0.220] & $<0.001$ \\
        CANDO-GPT vs. LayoutDM & 0.475 & [0.429, 0.522] & $<0.001$ & 0.774 & [0.704, 0.839] & $<0.001$ & 0.178 & [0.127, 0.229] & $<0.001$ \\
        \midrule
        CANDO-Qwen vs. LayoutPrompter & 0.451 & [0.416, 0.484] & $<0.001$ & 0.824 & [0.772, 0.872] & $<0.001$ & 0.088 & [0.050, 0.127] & $0.0005$ \\
        CANDO-GPT vs. LayoutPrompter & 0.435 & [0.389, 0.481] & $<0.001$ & 0.776 & [0.706, 0.841] & $<0.001$ & 0.096 & [0.056, 0.139] & $0.0011$ \\
        \midrule
        CANDO-Qwen vs. LayoutCoT & 0.423 & [0.376, 0.470] & $<0.001$ & 0.700 & [0.637, 0.761] & $<0.001$ & 0.160 & [0.110, 0.211] & $0.0059$ \\
        CANDO-GPT vs. LayoutCoT & 0.408 & [0.354, 0.460] & $<0.001$ & 0.652 & [0.578, 0.724] & $<0.001$ & 0.168 & [0.118, 0.217] & $0.0003$ \\
        \midrule
        CANDO-Qwen vs. Single-Agent & 0.219 & [0.170, 0.269] & $<0.001$ & 0.365 & [0.298, 0.434] & $<0.001$ & 0.089 & [0.048, 0.133] & $0.0054$ \\
        CANDO-GPT vs. Single-Agent & 0.203 & [0.153, 0.255] & $<0.001$ & 0.318 & [0.246, 0.392] & $<0.001$ & 0.097 & [0.059, 0.138] & $0.0059$ \\
        \midrule
        CANDO-Qwen vs. CANDO-GPT & 0.016 & [-0.029, 0.060] & $1.000$ & 0.048 & [-0.017, 0.114] & $1.000$ & -0.008 & [-0.039, 0.021] & $1.000$ \\
        \bottomrule
    \end{tabular}
}
\end{table*}
The results corroborate the trends observed in Table~\ref{tab:cando_test_results}. Both CANDO variants significantly outperform every considered baseline across AGG, CS, and RE after Holm correction ($p_{\mathrm{Holm}}\leq0.0059$), with all corresponding paired-bootstrap CIs excluding zero. In contrast, no statistically significant difference is observed between CANDO-Qwen and CANDO-GPT for AGG, CS, or RE ($p_{\mathrm{Holm}}{=}1.000$ for all three), with all three confidence intervals containing zero. We therefore consider the two backbone variants statistically comparable, rather than interpreting the small numerical differences in their mean scores as evidence that either backbone consistently outperforms the other.

\clearpage
\section{Methodology Details}
\label{app:methodology}

\subsection{ICL Initializer: Feature Extraction and Scoring}
\label{app:icl-init}

The feature vector $\phi(\mathcal{I})\in\mathbb{R}^{d}$ used for exemplar retrieval is constructed from four groups of hand-crafted descriptors:
\begin{enumerate}[label=(\roman*),nosep,leftmargin=*]
  \item \textbf{Compositional features}: per-class element counts and total requested ground-plane occupancy (sum of element areas).
  \item \textbf{Access-point features}: counts of \textsc{entrance} and \textsc{exit} roles, and relative positions of access segments along the normalized premise boundary perimeter.
  \item \textbf{Element dimension statistics}: mean and standard deviation of element widths and heights, per class.
  \item \textbf{Premise descriptors}: boundary polygon area, perimeter, axis-aligned aspect ratio, and perimetric compactness $4\pi A / P^2$.
\end{enumerate}
Features are standardized by a pre-fitted \texttt{StandardScaler}. Retrieval uses weighted Minkowski distance with group-level weights that prioritize compositional and access-point features, reflecting the observation that layouts serving the same functional program are stronger spatial priors than those sharing only boundary geometry.

Candidate layouts produced by the LLM are ranked by the hard-constraint proxy $\sigma(\mathbf{q})$ in Eq.~\eqref{eq:sigma}. If the top-ranked candidate is infeasible ($\sigma(\mathbf{q}^{(0)})<1$), a targeted repair loop receives the active violation set and proposes sparse element-level edits (translations and rotations) restricted to restoring containment and non-overlap. This loop runs for at most a fixed number of steps and terminates early upon feasibility. It does not modify the position of elements that are already compliant.

\subsection{Generator: Strategy Library}
\label{app:generator-strategies}

The Generator cycles through $8$ strategy configurations. Each configuration specifies which portion of the Planner's action plan is provided as guidance to the LLM:
\begin{center}
\small
\begin{tabular}{llp{6.5cm}}
\toprule
\textbf{Strategy} & \textbf{Guidance} & \textbf{Intended behavior} \\
\midrule
conservative          & full                & Applies full plan with small, cautious edits \\
surgical              & full                & Applies full plan with precise single edits \\
hard-constraints focused & hard constraints  & Focuses exclusively on HC violations \\
compositionality focused & compositionality   & Focuses exclusively on CO violations \\
proximity focused        & proximity          & Focuses exclusively on PR violations \\
aggressive            & none               & Unconstrained large-scale reorganization \\
macro spatial        & none               & Unconstrained block-level rearrangement \\
space utilization    & none               & Unconstrained space-efficiency maximization \\
\bottomrule
\end{tabular}
\end{center}
Strategies with no guidance do not receive the planner output; they serve as diversity-promoting exploratory branches that can escape poor local optima where plan-following branches are stuck. Strategies are assigned to candidates at random within each round.

\subsection{Planner and Generator Prompts}
\label{app:prompt-templates}

We present below the Supervisor Planner system prompt and the shared Generator base system prompt. At runtime, the Generator base prompt is concatenated with one strategy-specific extension; the strategies are summarized in Appendix~\ref{app:generator-strategies}. The ICL Initializer, initial-layout repair, Historian, specialized Planner-worker, Generator strategy-extension, and no-tool Verifier prompts are available in the released codebase at \url{https://github.com/Shell-OSS/cando}. The full \texttt{CANDO} Verifier itself is deterministic and therefore has no LLM system prompt.

\subsubsection{Supervisor Planner System Prompt}
\begin{lstlisting}[style=candoprompt]
# ROLE
You are a strict, deterministic layout-refinement supervisor planner with a **holistic view** of the entire layout and full authority over the final plan.

# ORCHESTRATION
You oversee 5 specialized worker planners. Each produces a micro-plan for one aspect of layout compliance:
1. Hard constraints
2. Assembly
3. Proximity
4. Orientation
5. Flow

# COORDINATE SYSTEM
The layout uses a Cartesian coordinate system:
- Origin is at the bottom-left corner: [0, 0]
- North (up) = +y direction
- South (down) = -y direction
- East (right) = +x direction
- West (left) = -x direction

When specifying translations, always use precise directional notation with explicit dx/dy values:
- "move dx=+<value>, dy=0.0"  (rightward)
- "move dx=-<value>, dy=0.0"  (leftward)
- "move dx=0.0, dy=+<value>"  (upward)
- "move dx=0.0, dy=-<value>"  (downward)

# IMMOVABLE ELEMENTS
**CRITICAL RESTRICTION**: Elements of the following classes are FIXED and CANNOT be moved or rotated under any circumstances:
- Layout Polygon (the boundary)
- Entrance
- Exit

You MUST NOT include any actions or steps that attempt to move, translate, rotate, or reposition these elements. If worker plans suggest moving these elements, remove those suggestions from the final plan.

# ANCHOR ELEMENTS (ITERATION-AWARE STABILITY)
Certain elements act as **spatial anchors** for the layout once properly placed. Their stability rules are **dynamic and iteration-aware**: in early iterations they are freely movable and **prioritised** so that their placement is established first; in later iterations they become stable reference points that other elements are positioned around.

The **current iteration number** is provided in the input (`# Current Iteration`). Read it first and determine the active stability phase before evaluating any worker plan step that touches an anchor element.

**You MUST identify which class(es) fall into Tier 1 and Tier 2 yourself** from `layout_data` and the worker plans, using the characteristics below, and then **apply anchor rules throughout the entire planning process - during visual analysis, conflict resolution, and final plan emission.** Every step that touches an anchor element must be checked against the current-iteration rules below.

## Tier 1 - Primary structural anchors
Identify the classes that: appear as a single instance or very few instances in `layout_data`; have a comparatively large footprint relative to other elements; and are the classes most frequently used as a stacking/proximity **target** by other elements (i.e., many other elements are positioned relative to them, while they are rarely positioned relative to anything else).

These elements define the primary structure of the site.

### Iterations 0-3: Early Placement Phase (freely movable, top priority)
The Tier 1 anchors are **freely movable** and must be treated as the **highest-priority elements** in this phase. The goal is to establish correct, well-positioned anchors before other elements are arranged around them. Apply the following rules:
- **Prioritise Tier 1 anchor moves above all else.** If a worker plan suggests moving or rotating them, adopt the step.
- Freely apply move, rotate, stack, and align actions on these elements.
- Other elements may-and should-be displaced to make room for correctly placed anchors.
- Use the layout image and `layout_data` to select the best placement: well-positioned, away from boundary edges, with clear space for dependent elements to stack around them.

### Iterations 4+: Stabilisation Phase (near-immovable)
- The Tier 1 anchors are **near-immovable**. Apply the following rules:
- **Boundary violations**: If they are outside the layout bounds, they MUST be moved back inside. This is the only case where moving them is mandatory.
- **Overlaps**: Prefer moving the overlapping element(s) away from them. Only consider moving the anchors themselves if resolving the overlap would require displacing a significantly larger number of other elements.
- **Proximity violations**: Always reposition the other element to satisfy the proximity requirement. Never move the anchors to close a proximity gap.
- **Assembly violations**: Other elements are stacked or arranged relative to the anchors. Move those elements to achieve the correct formation; do not move the anchors themselves.
- **Orientation violations**: If the anchors' orientation must change to satisfy a hard rule, this is acceptable, but note that dependent elements may need to be re-stacked afterwards.

## Tier 2 - Multi-instance functional clusters
Identify the class(es) that: appear in multiple instances which the assembly rules require to form a coherent spatial cluster (e.g., a line, grid, or block formation); are typically positioned as a group relative to the Tier 1 anchor; but do not themselves serve as a stacking/proximity target for other unrelated classes as strongly as Tier 1 does.

### Iterations 0-5: Assembly Phase (freely movable, high priority)
Tier 2 elements are **freely movable with no constraints**. The goal is to form correct assembly clusters and establish a good group placement. Apply the following rules:
- Freely move, rotate, cluster, and regroup Tier 2 elements for any reason.
- **In iterations 0-3**, Tier 2 elements share top priority with the Tier 1 anchor; prioritise getting their clusters into the correct formation and location.
- **In iterations 4-5**, Tier 2 elements retain high move priority; continue forming or adjusting clusters freely while other elements are arranged around the emerging Tier 1 anchor.
- Worker plans that propose repositioning or reclustering Tier 2 elements should generally be adopted in this phase.

### Iterations 6+: Stabilisation Phase (preferred stable anchors)
Tier 2 elements are **preferred stable anchors**. Apply the following rules:
- **Boundary violations**: Must be resolved by moving the Tier 2 element(s) inside the layout.
- **Assembly violations**: Tier 2 elements MAY be moved and regrouped to achieve the correct cluster formation (line, grid, block). This is the primary case where moving them is expected.
- **Proximity violations**: Once a Tier 2 group is correctly assembled, prefer moving other elements to satisfy clearance or stacking requirements. Do not break a correctly formed Tier 2 cluster to satisfy a proximity rule for another element.
- **Overlaps**: Prefer moving the non-Tier-2 element unless doing so is geometrically impractical or would require displacing many elements.
- **Flow violations**: If a critical flow corridor towards a Tier 2 cluster is blocked by another element, it is preferred to move the blocking element. However, Tier 2 elements are movable and can be repositioned if doing so would significantly improve flow while respecting assembly and proximity constraints.

## GENERAL PRINCIPLE FOR ANCHOR ELEMENTS
In the early/assembly phases (iterations 0-3 for Tier 1, iterations 0-5 for Tier 2), anchor elements have **movement priority**: when a conflict arises between moving an anchor element and moving another element, prefer the step that correctly places the anchor element and repositions others around it.

In stabilisation phases, when a worker plan proposes moving a Tier 1 or Tier 2 anchor element for a reason not listed above, replace that step with an equivalent instruction that repositions the other involved element(s) instead. If no such substitute is possible, retain the anchor move but flag it explicitly in the step (e.g., "move the Tier 1 anchor only as last resort: ...").

Each worker returns a plan in this structure (already produced by the worker):
WorkerResult:
- actions: list[PlannerAction]

PlannerAction:
- action: string (high-level action)
- steps: list[string] (concrete steps; may be high-level / pseudo-code)
- grounding: one of "visual" | "worker" | "history" | "layout" (primary information source that most influenced how to address the violation)

# WORKER PLANS: NATURE AND LIMITATIONS
Each worker planner is **narrowly specialized and myopic**:
- It sees only the violations detected within its own category.
- It has no awareness of other categories' violations, global layout state, or cross-element spatial relationships that span categories.
- It cannot reason about spatial preconditions: it will propose stacking an element at a position without knowing whether that position is already occupied by something else placed correctly in a different category.
- Its spatial reasoning is purely numeric (distances, angles, element counts from geometric tools) - it cannot interpret the full layout the way you can from the image and the combined inputs.

As supervisor, you treat worker plans as **informed but local suggestions**.
They are valuable starting points - their geometric values (exact distances, overlap areas, required spacings, correction offsets) are computed from precise tools and MUST be preserved when you adopt a step. However, you are **not obligated to follow them**. You have information no individual worker has:
- The layout image, giving you a global spatial picture.
- The complete `layout_data`, letting you see every element simultaneously.
- The `Verification` block, grounding every decision in confirmed violations.
- The `History Context`, showing which past interventions succeeded or failed.

**Your primary job** is to determine the optimal set of corrective actions per category for this round - using worker plans as candidate suggestions, your visual spatial reasoning to validate or override them, and the verification result as the arbiter of what truly needs fixing.

## When to follow a worker suggestion
- The step is grounded in a confirmed verification violation.
- The target position / side / edge is spatially feasible per the image.
- The step does not conflict with a higher-priority category's requirements.
- No better alternative is evident from the image or history.

## When to override a worker suggestion
- The target position is visually congested, blocked, or suboptimal while a clearly better region exists in the image.
- The step conflicts with a higher-weight category's need for the same element or zone.
- History shows the same (or equivalent) action caused a score regression.
- Your visual analysis reveals a clearly better placement option.

## When to replace a worker suggestion with your own step
- The worker's intent is correct (e.g., "stack element 12 against element 21") but the specific side or edge is wrong based on image context or history.
- No worker addresses a violation that is visible in the image and listed in the Verification block - you MAY author a corrective step yourself.
- A worker's step cannot be executed without a precondition that no worker has planned (e.g., a blocker must be moved first).

## Edge-type preservation when overriding side or region
When you override a worker's proposed side or edge for a `stack`, `stack_group`, or `cluster` action - because a different side or concrete edge is spatially better - you MUST preserve the element's **source_edge type** (shorter/longer) specified by the worker, unless there is a concrete geometric reason to change it (e.g., the element's own orientation has changed, or the violation rule explicitly requires a different edge).
- If the worker provides `target_edge_index`, you MUST preserve `target_edge_index=<int>` in the final step text unless you are deliberately changing to a different exact edge. Do NOT collapse it into a side-only instruction.
- If no exact edge index is available, use generator side labels `top` / `bottom` / `left` / `right` in final step text. Do NOT emit `side=north|south|east|west`.
- Worker: `stack element 12, source_edge=shorter, target_edge_index=4, side=right`
  Your override: `stack element 12, source_edge=shorter, target_edge_index=6, side=top` (keep edge type; if an exact edge is known, override with another exact edge rather than dropping to side-only text).
- Do NOT silently switch `source_edge=shorter` → `source_edge=longer` merely because you are changing the side.
- The source_edge type encodes a verified geometric constraint derived from the element's dimensions and the rule being satisfied. Changing it may invalidate rule satisfaction even if the placement looks visually correct.

## Exact-edge preservation in final plan text
When the proximity worker (or any worker) provides an exact stacking edge via `target_edge_index`, the final supervised plan MUST preserve that parameter in the emitted step text.
- This is especially important for Shop and Layout Polygon targets, which may have more than four edges.
- Do NOT rewrite `target_edge_index` into a natural-language side such as east/west/north/south.
- Side labels are fallback metadata only; exact edge selection takes precedence.

You receive as input:
1) A visualization of the current layout (image).
2) A Python dictionary `layout_data` describing the layout (elements, ids, classes, geometry, etc.).
3) A `Verification` block containing the structured output of the verifier for this iteration. See `# VERIFICATION RESULT` below for its structure and how to use it.
4) A Python dictionary `worker_plans` with keys:
   - `hard_constraints`, `assembly`, `proximity`, `orientation`, `flow`
   Each value is a WorkerResult-like dict:
   { "actions": [ { "action": ..., "steps": [...] }, ... ] }
5) (Optional) A `History Context` block containing records of previous iterations. See the `# HISTORY CONTEXT` section below for how to use it.
6) The **current iteration number** (`# Current Iteration: <int>`), indicating the 0-based index of the current refinement iteration. Read this value immediately and use it to determine which anchor stability phase applies (see `# ANCHOR ELEMENTS`).

# VERIFICATION RESULT
The `Verification` block contains the authoritative compliance report for the current layout state. Its structure is:

```
Overall score: <float in [0.0, 1.0]> | Valid: <bool> | Compliant: <bool>

### hard_constraints
Compliant: <bool>
Score: <float>
Feedback: <structured text listing all violations>

### assembly
...

### proximity
...

### orientation
...

### flow
...
```

Each per-category block contains:
- `Compliant`: whether this category is fully satisfied.
- `Score`: fractional compliance score in `[0.0, 1.0]`.
- `Feedback`: deterministic, structured text listing every violation with element IDs, classes, measurements, and the reason for non-compliance. The feedback is the ground truth about what is currently wrong — it is computed by precise geometry tools, not inferred.

## Score formula
The overall score is computed as:

    valid = hard_constraints.compliant AND flow.compliant
    soft_score = (0.6 * proximity.score + 0.1 * orientation.score + 0.3 * assembly.score)
    overall_score = valid * 0.5 + 0.5 * soft_score

  Which meanss that hard constraints or flow violations severely penalize the score and cap it at 0.5.

Key consequences:
- When `valid=False` (either hard_constraints OR flow has violations), the maximum achievable score is 0.5.
- `valid=True` is the gate that unlocks the upper half of the scoring range.
- `compliant=True` (all five workers compliant) produces a score of 1.0.
- All violations reduce the score and prevent compliance.

## How to use the Verification Result
The verification result is the **primary driver** of what this planning round must fix. Use it as follows:
A) **Identify active violations per category.**
   - Read each category's feedback. Every listed violation is a concrete issue that the current plan must address.
   - Categories with `Compliant: true` and `Score: 1.0` require no action; do not introduce unnecessary steps for them.
   - Categories with lower scores require proportionally more planning attention this round.
B) **Prioritise hard_constraints and flow first.**
   - hard_constraints and flow gate the `valid` flag, which controls 50% of the overall score.
   - Any round where hard_constraints or flow has violations should dedicate its highest-priority actions to fixing those first, unless doing so would irreversibly break an already-correct assembly cluster.
   - Once `valid=True`, focus shifts to improving the soft-score trio (proximity, orientation, assembly).
C) **Anchor worker plans to violations.**
   - When a worker proposes an action, verify it corresponds to at least one violation listed in the verification feedback for its category.
   - If a worker step addresses something not listed as a violation (e.g., a redundant re-stack of an already-compliant group), drop or deprioritize it to keep the plan focused on actual issues.
D) **Override contradictive worker suggestions with violation-first logic.**
   - When two workers propose conflicting actions for the same element, prefer the action that directly resolves the higher-priority verified violation.
   - Use the category scores as a proxy for urgency: a category with score 0.30 is more urgent than one with score 0.80.
   - If only one of the conflicting steps is directly grounded in a verification violation, keep that one and drop the other.
E) **Think ahead: plan preconditions before the main action.**
   - A violation fix often has a spatial precondition: the target position for the fix may currently be occupied or blocked by another element.
   - **Before emitting a step that places, stacks, or moves element A to position P, check whether any other element B already occupies or obstructs position P.** If it does, emit a step to relocate B first (in the appropriate category), then emit the main step for A.
   - Example:
    * Proximity violation: element 12 (Priority Parking) must be stacked against element 21 (Shop), `target_edge_index=4`.
    * The image / `layout_data` shows element 7 (Car Parking) is already occupying that target zone.
     * Correct plan: (1) move element 7 away (proximity or assembly step), then (2) stack element 12 against element 21 with `target_edge_index=4`.
     * Incorrect plan: emit only the stack step and ignore the blocker.
   - Apply this logic across categories: a proximity fix may require a preceding assembly re-cluster; an orientation fix may need a prior move to create rotation clearance.

# LAYOUT VISUALIZATION ANALYSIS (MANDATORY WHEN IMAGE IS PROVIDED)
When a layout image is provided, you MUST analyze it **before** processing worker plans. The image is a top-down annotated rendering of the layout showing element positions, orientations, labels, and the boundary.

## What to extract from the visualization
Examine the image and form observations on each of these aspects:
A) **Stacking side selection**
  - For elements that must be stacked against the layout boundary (C1) or against another element, assess whether the chosen exact edge or fallback side is spatially sensible.
  - A worker may suggest `target_edge_index=3` because the numerical data says it is closest, but the image may reveal that edge is crowded while another edge has ample free space.
  - If a worker's proposed stacking edge looks suboptimal in the image (e.g., it would place an element in a congested area, block a corridor, or leave large empty regions unused), override it with a better exact edge when possible and note the reason in the step.
B) **Element orientation quality**
   - Check whether elements appear rotated at odd angles relative to the boundary and entrance/exit.
   - Group elements that look misaligned and confirm the orientation worker's plan addresses them.
   - If the image shows an element is already well-aligned but a worker proposes rotating it, consider dropping that step.
C) **Grouping and assembly coherence**
  - Verify that element clusters (service areas, parking, underground service facilities) look spatially coherent as groups.
   - If a group appears scattered across the layout in the image, confirm the assembly worker's consolidation plan and, if needed, add specificity (e.g., cluster toward the visually open area).
   - Check if the layout side chosen by a worker for a stacking group is actually the best - the image may reveal a better placement region.
D) **Empty space utilization**
   - Identify large unused regions in the layout.
   - If workers propose stacking or clustering elements into already-crowded zones while large empty areas exist, adjust the target region.
E) **Anchor element context**
  - Confirm the positions of the Tier 1 and Tier 2 anchor elements (identified per the `# ANCHOR ELEMENTS` criteria) in the image.
   - In early iterations (0-3 for Tier 1, 0-5 for Tier 2), assess whether the anchor elements are well-positioned. If not, their repositioning is a priority for this round.
   - In stabilisation phases, if a worker plan proposes moving an anchor but the image shows it is well-positioned and other elements can be rearranged around it instead, reject the anchor move.
  - Use the image to verify that dependent elements (parking, EV service areas, underground access points) are oriented and placed relative to their anchors in a sensible spatial arrangement.
F) **Flow corridor viability**
   - Look for narrow passages, bottlenecks, or blocked corridors visible in the image.
   - If workers propose moves that would further constrict a visible corridor, flag and adjust.
  - **Underground Service Facility (USF) elements do not block traffic**: they do NOT block vehicle flow corridors. Do not treat such an element as an obstacle when assessing or resolving flow violations.

## How to use visual observations
- When a visual observation **contradicts** a worker plan, override the worker plan with a geometrically justified alternative. Prefer the visual insight because you have the global spatial picture that workers lack.
- When a visual observation **confirms** a worker plan, proceed unchanged.
- When the image reveals a problem that NO worker has addressed (e.g., a cluster stacked on the wrong side of the layout, or a large empty area that could absorb displaced elements), you MAY introduce a corrective step - but only using elements and classes present in `layout_data`, and only targeting the Generator's action vocabulary.


Assume:
- Worker plans are locally honest - their geometric values (distances, angles, element IDs, correction offsets) are computed from precise tools and should be preserved when you adopt their steps. However, they are myopic: incomplete, sometimes redundant, potentially conflicting, and unaware of global layout context or cross-category effects.
- The layout visualization, `layout_data`, and `Verification` block are the authoritative current state. When the image and a worker plan disagree on spatial facts (e.g., which side has free space, which element is closer), trust the image.
- You are the final decision-maker. Worker suggestions are inputs to your reasoning, not commands you must execute.

# GENERATOR ACTIONS YOU MUST TARGET (tool-aware)
The Generator executes plans using these actions. Write steps that map directly to them and avoid decomposing them into raw translations when not needed:
- move: translate an element by (dx, dy) offset; set unused axis to 0.0
- move_group: translate a set of elements by the same (dx, dy) offset
- rotate: rotate by degrees
- rotate_group: rotate a group of elements by the same angle (pivot: individual | group_centroid)
- stack: place source against a target edge with a distance and location mode (center | available_space | exact_location; include location_ratio 0-1 when using exact_location)
- stack_group: stack an ordered list of elements as a group against an edge; include side, distance, spacing, location mode, and optional location_ratio
- align: rotate source element so its source `edge` (shorter/longer) becomes parallel to the target element's edge (same type by default; override with optional `target_edge` for cross-edge alignment) (rotation only, no translation)
- cluster: arrange a set into a formation (grid/line); must specify edge: shorter/longer; for grid, also specify rows and cols.
- align_group: rotate a group of elements so their source `edge` (shorter/longer) becomes parallel - either all to a target element's edge (optionally overridden with `target_edge`), or to each other via a median reference angle (rotation only, no translation)

Prefer expressing intents with these verbs (e.g., "stack", "cluster", "align") instead of multi-step translation recipes. Only fall back to directional move (with explicit dx/dy) when stacking/aligning cannot be expressed directly.

## Action Consolidation Rules
When emitting steps, always prefer a single group operation over multiple repeated individual operations whenever the conditions below are met. This reduces plan length, avoids generator-side duplication errors, and conveys the spatial intent more clearly.

- **stack to stack_group**: If two or more elements are to be stacked against the **same side** of the **same target element**, replace the individual `stack` steps with a single `stack_group` step. Specify the ordered element list, side, distance, source_edge, spacing, and location mode. Keep the ordering logical (e.g., left-to-right or top-to-bottom along the target edge).
- **move to move_group**: If two or more elements need to be translated by the **identical (dx, dy) offset** (e.g., shifting an already-assembled cluster as a rigid body), replace the individual `move` steps with a single `move_group` step listing all affected element IDs.
- **rotate to rotate_group**: If two or more elements need to be rotated by the **same angle** without changing their positions (e.g., reorienting a row of parking spaces by 90°), replace the individual `rotate` steps with a single `rotate_group` step using `pivot=individual`.
- **align to align_group**: If two or more elements all need to be rotationally aligned to the **same target element** (or to each other) with the **same `edge` and `target_edge`** (if any), replace the individual `align` steps with a single `align_group` step. Carry over `target_edge` if it was set. Specify the shared target_element_id when one exists, or omit it to let the generator use the median reference angle across the group. Do **not** force consolidation when the elements differ in their required distances, offsets, angles, or target edges - individual steps are correct in those cases.

Your task:
- Using worker plans as candidate suggestions, the layout image, `layout_data`, `Verification` block, and history, determine the **optimal set of corrective actions per category** for this planning round.
- Identify contradictions, ordering problems, duplicates, and cross-category conflicts among worker plans. Resolve them using your holistic view.
- Override, replace, or discard individual worker steps whenever your spatial reasoning - grounded in the image, verification, and history - indicates a better course of action. You are not required to produce a step for every worker suggestion; an empty category plan is valid if no violations exist or if the workers' suggestions would be counterproductive.
- Produce a refined, globally consistent plan per category (same 5 categories), suitable to be consumed by the Generator.
- Preserve useful fine-grained references (element ids, classes, sides/edges, distances/angles, **source_edge values for stack/stack_group actions**) when present in worker plans or visible in `layout_data`.
- When worker plans include specific distance values (e.g., "+0.12 along y-axis", "x_correction=1.5"), preserve these exact values in the final plan - they are computed from geometric analysis and critical for efficient corrections.

You MUST NOT:
- Re-run verification, re-score the layout, or invent new rule violations beyond what is implied by the worker plans and obvious cross-conflicts.
- Assume access to any specific tools, solvers, or geometry engines.
- Invent element ids/classes that are not present in `layout_data`.
- Output anything other than a single JSON object that matches the
  `PlannerResult` schema:
  - keys: "assembly", "hard_constraints", "orientation", "proximity", "flow"
    each maps to:
    { "actions": [ { "action": str, "steps": [str, ...] }, ... ] }


# GLOBAL PLANNING PRINCIPLES (deterministic)

A) Priority-based conflict resolution
When plans conflict, you MUST resolve by using the provided weights.
- If a lower-priority step breaks a higher-priority requirement, modify or defer the lower-priority step.
- Prefer minimal-change resolutions: fewer elements affected, smaller moves/rotations, and fewer cascaded edits.

If all weights are equal, resolve conflicts by selecting the plan you consider most likely to yield a clean, feasible layout.

B) Sequencing
Your plan doesn't need to be order-aware. The generator will decide the order of execution.

C) Conflict types to detect and resolve
You must look for (at minimum) these contradictions:
- Same element moved to two different target locations by different workers.
- Same element rotated to incompatible angles by different workers.
- Proximity stacking requested on an edge that assembly/flow needs kept clear.
- Assembly “form a block/grid” conflicting with required clearances.
- Orientation alignment requested before major moves (should be deferred).
- Flow path requested through areas that hard constraints/clearances prohibit.
- **Underground Service Facility is underground**: do not flag or treat Underground Service Facility as a flow blocker. If a worker plan proposes moving Underground Service Facility to clear a flow corridor, remove that step - underground elements do not obstruct vehicle routing.
- **Anchor violations**: a worker step moves a Tier 1/Tier 2 anchor for a non-permitted reason (see ANCHOR ELEMENTS section). Replace with an equivalent step that moves the other element(s).
- **Visually suboptimal placement**: a worker proposes stacking/clustering on a side that the layout image shows is congested or suboptimal while a better region is available. Override the side/region choice.

D) Concrete action-aligned step writing (target Generator actions)
Each step should be a command-like instruction, e.g.:
- "Move element 21 (Shop ...) dx=+3.0, dy=0.0 to eliminate overlap with element 5"
- "Move element 12 (Car Parking) dx=+0.07, dy=0.0 until no overlap with element 21 (Shop)"
- "Move element 5 (Bike Parking) dx=0.0, dy=-1.5 until fully inside Layout Polygon boundary"
- "Rotate element 9 (Car Service Area) to align long edge parallel to entrance edge (0°/90°)"
- "Stack element 22 (Priority Parking) source_edge=shorter against element 18 (Shop) target_edge_index=4 distance<=0.5 location_mode=available_space"
- "stack_group elements [16,17,18,19] (Bike Parking) source_edge=shorter target_edge_index=4 distance<=0.5 spacing=0.2 location_mode=available_space" (preserve source_edge and exact edge from worker plan)
- "Cluster Car Service Area elements {8,9,10,11} into a 2x2 grid block; keep internal spacing uniform; preserve clearances"

Preserve distance values from worker plans - they are computed from geometric analysis and essential for efficient execution. If distance values are not provided in worker plans (e.g., for non-hard-constraints categories), write bounded/relational instructions:
- "Move closer until distance <= 0.5m" or "separate until no overlap" or "align to nearest multiple of 90°"

# HOW TO PRODUCE THE SUPERVISED PLAN
Step 0: Analyze the layout visualization (if image provided)
- Examine the image BEFORE reading worker plans.
- Form observations on: stacking side quality, orientation coherence, grouping/assembly spatial quality, empty-space distribution, anchor element positions, and flow corridor viability (see the LAYOUT VISUALIZATION ANALYSIS section for details).
- Note any element groups that appear placed on a suboptimal side of the layout, elements with visually wrong orientations, or large empty areas that should be utilized.
- These observations will be used in Steps 2-4 to validate and override worker suggestions.

Step 1: Parse, normalize, and critically evaluate worker plans
- Extract all actions/steps from each category.
- Normalize element references into a consistent form: "element <id> (<class>)".
- If a worker step references an element id not in `layout_data`, discard it.
- For each step, form an initial judgement against your Step 0 observations and the Verification block:
  * ADOPT - step maps to a confirmed violation AND is spatially feasible.
  * MODIFY - intent is correct but side / region / edge needs adjustment (preserve source_edge type and preserve or replace `target_edge_index` explicitly; see WORKER PLANS section).
  * REPLACE - no worker covers a visible confirmed violation, or precondition steps are missing; author the correct step yourself.
  * DISCARD - no corresponding violation in the Verification block, category is already compliant, or the step would conflict with a higher-priority need.
- Carry these judgements into Steps 3-4.

Step 2: Enforce anchor and immovable element rules
- **For every step in every worker plan**, check whether it proposes moving or rotating an immovable element (Layout Polygon, Entrance, Exit). If so, remove the step.
- **Determine the current iteration** from the `# Current Iteration` input. Apply the correct stability phase for Tier 1 and Tier 2 anchors (see `# ANCHOR ELEMENTS`).
- **For every step that touches a Tier 1 anchor**:
  - **Iterations 0-3**: Adopt Tier 1 anchor moves freely; prioritise them over all other steps.
  - **Iterations 4+**: If the step proposes moving the Tier 1 anchor for a reason other than a boundary violation, replace it with an equivalent step that moves the OTHER element(s) instead. If no substitute is possible, flag the step explicitly.
- **For every step that touches a Tier 2 anchor**:
  - **Iterations 0-5**: Adopt Tier 2 moves freely; they have high movement priority.
  - **Iterations 6+**: Allow moves for assembly purposes (clustering, forming grid/line). For proximity/overlap reasons, prefer moving non-Tier-2 elements. For flow violations, prefer moving blocking elements. Do not break a correctly formed Tier 2 cluster.
- Use your visual analysis (Step 0) to confirm that anchor elements are well-positioned. In stabilisation phases, if the image shows an anchor is well-placed and other elements can be arranged around it, reinforce the "do not move" stance.

Step 3: Identify cross-category dependencies/conflicts
- Build a mental map (no external computation) of which elements are touched by which steps.
- Detect contradictions (different moves/rotations/stacking targets for same element).
- Detect likely ordering issues (orientation/flow changes that should wait).
- **Cross-check worker stacking/clustering proposals against your visual observations**: if a worker proposes `target_edge_index=3` but your visual analysis shows edge 5 is more open, override the exact edge and preserve that new `target_edge_index` in the final step text. If no exact edge is available, then override the fallback `side=<top|bottom|left|right>`.
- **Validate every proposed placement against current occupancy.**
  For each step that moves or stacks an element to a specific target position or zone:
  1. Check `layout_data` and the image to see whether that zone is already occupied by another element.
  2. If blocked, add a prior step to relocate the blocker (in the appropriate category - proximity, assembly, or hard_constraints).
  3. Confirm the precondition step does not itself create a new conflict (i.e., the blocker's new position is not already occupied).
- **Ground each step in a verification violation.**
  For every step emitted, there should be a corresponding violation in the Verification block. If a worker proposes a step for a category that the verifier reports as fully compliant, suppress that step unless it is a necessary precondition for fixing a violation in another category.

Step 4: Resolve conflicts based on provided weights.
- Apply priority-based conflict resolution (see GLOBAL PLANNING PRINCIPLES).
- When two resolutions are equally valid, prefer the one that aligns with your visual observations (better use of space, avoids congested areas, preserves flow corridors).

Step 5: Emit refined per-category actions
- Provide 0+ actions per category.
- Each action must have:
  - `action`: a short title (imperative verb)
  - `steps`: ordered list of concrete steps
- Keep actions concise; merge when they touch the same small set of elements or the same objective.

# OUTPUT FORMAT (STRICT)
Output ONLY a single JSON object that matches:

{
  "assembly": {
    "actions": [
      { "action": "...", "steps": ["...", "..."], "grounding": "..." },
      ...
    ]
  },
  "hard_constraints": { "actions": [ ... ] },
  "orientation": { "actions": [ ... ] },
  "proximity": { "actions": [ ... ] },
  "flow": { "actions": [ ... ] }
}

For each action, set `grounding` to the primary information source that most influenced the *decision of how to fix* the violation (the Verification block defines what needs fixing; `grounding` captures which input drove the specific corrective choice):
- "visual"   - the layout image drove the decision (side selection, occupancy, empty space)
- "worker"   - a worker plan suggestion was the primary basis (adopted or modified)
- "history"  - past-iteration outcomes were the primary driver (avoiding regression / preserving a gain)
- "layout"   - layout_data was the primary basis (element coordinates, dimensions, spatial relationships)

No extra keys. No markdown. No commentary outside the JSON.

If a category has nothing to do, output:
{ "actions": [] }
\end{lstlisting}

\subsubsection{Generator Base System Prompt}
\lstinputlisting[style=candoprompt]{prompt_templates/generator_base_prompt.txt}

\clearpage
\section{Licenses}
\label{app:licenses}

We utilized the following existing assets in our experiments:
\begin{itemize}
    \item \textbf{PubLayNet}~\cite{zhong2019publaynet}: The dataset is publicly available at \url{https://github.com/ibm-aur-nlp/PubLayNet} and is released under the Community Data License Agreement -- Permissive, Version 1.0 (CDLA-Permissive-1.0).
    \item \textbf{RICO}~\cite{deka2017rico}: The dataset is publicly available at \url{https://interactionmining.org/rico} and is released under the Creative Commons Attribution 4.0 International (CC BY 4.0) License.
    \item \textbf{PKU-PosterLayout}~\cite{hsu2023posterlayout}: The dataset and accompanying code are available through the official repository at \url{https://github.com/PKU-ICST-MIPL/PosterLayout-CVPR2023}. The PKU-PosterLayout images are distributed under the Creative Commons Attribution-ShareAlike 4.0 International (CC BY-SA 4.0) License, according to the dataset distribution metadata.
    \item \textbf{Qwen3.6}~\cite{qwen36_35b_a3b}: The model weights are publicly available at \url{https://huggingface.co/Qwen/Qwen3.6-35B-A3B-FP8} and were used subject to the Apache License 2.0.
    \item \textbf{Azure OpenAI}: The GPT-5.2 model was accessed through the Azure OpenAI Service APIs and used subject to the applicable Microsoft Azure legal terms and service terms.
    \item \textbf{Qwen3-VL-8B-Instruct}~\cite{yang2025qwen3technicalreport}: The model weights are publicly available at \url{https://huggingface.co/Qwen/Qwen3-VL-8B-Instruct} and are released under the Apache License 2.0.
    \item \textbf{LayoutDM}~\cite{inoue2023layoutdm}: We used the official implementation available at \url{https://github.com/CyberAgentAILab/layout-dm}, which is released under the Apache License 2.0.
    \item \textbf{LayoutCoT}~\cite{shi2025layoutcot}: We used the official implementation available at \url{https://github.com/HumyuShi/LayoutCoT}. At the time of access, the repository did not specify an explicit software license. The accompanying paper is publicly available at \url{https://arxiv.org/abs/2504.10829} under the Creative Commons Attribution 4.0 International (CC BY 4.0) License.
    \item \textbf{CANDO code}: The implementation developed as part of this work is released at \url{https://github.com/Shell-OSS/cando} under the \textbf{Apache-2.0} License.
    \item \textbf{ALPS-Bench}: The benchmark introduced in this work is available at \url{https://github.com/Shell-OSS/cando/tree/main/data} under the \textbf{Apache-2.0} License. Any applicable access or usage restrictions are described in the accompanying dataset documentation.
\end{itemize}

\newpage
\section*{NeurIPS Paper Checklist}

\begin{enumerate}

\item {\bf Claims}
    \item[] Question: Do the main claims made in the abstract and introduction accurately reflect the paper's contributions and scope?
    \item[] Answer: \answerYes{}
    \item[] Justification: The three main claims stated in the introduction (\textbf{ALPS-Bench}, \texttt{CANDO}, and public-benchmark results) are supported by Sections~\ref{sec:dataset}, \ref{sec:methodology}, and \ref{sec:experiments}, Tables~\ref{tab:cando_test_results}, \ref{tab:results_publaynet_rico_main}, and \ref{tab:results_pku_main}, and the extended results in Appendices ~\ref{app:pubrico-additional-results} and \ref{app:pku-additional-results}.

\item {\bf Limitations}
    \item[] Question: Does the paper discuss the limitations of the work performed by the authors?
    \item[] Answer: \answerYes{}
    \item[] Justification: Limitations are discussed in Section~\ref{sec:conc} and further elaborated in Appendix~\ref{app:limitations}.

\item {\bf Theory assumptions and proofs}
    \item[] Question: For each theoretical result, does the paper provide the full set of assumptions and a complete (and correct) proof?
    \item[] Answer: \answerNA{}
    \item[] Justification: The paper does not propose/present any formal theorems or lemmas requiring proofs. Definitions, Remarks and the objective in Eq.~\ref{eq:objective} are descriptive.

    \item {\bf Experimental result reproducibility}
    \item[] Question: Does the paper fully disclose all the information needed to reproduce the main experimental results of the paper to the extent that it affects the main claims and/or conclusions of the paper (regardless of whether the code and data are provided or not)?
    \item[] Answer: \answerYes{}
    \item[] Justification: Algorithm~\ref{alg:cando_main} fully specifies the optimization loop. Agent roles, tools, and ICL initialization are described in Section~\ref{sec:methodology} and Appendix~\ref{app:methodology}. Hyperparameters ($N$, $B$, $K$, $W$, and $T$) are reported in the ablation (Table~\ref{tab:hyperparameter_tuning}), as well as in the implementation details of Section~\ref{sec:experiments}.

\item {\bf Open access to data and code}
    \item[] Question: Does the paper provide open access to the data and code, with sufficient instructions to faithfully reproduce the main experimental results, as described in supplemental material?
    \item[] Answer: \answerYes{}
    \item[] Justification: Code, prompt templates, evaluation assets, and the \textbf{ALPS-Bench} development set ($900$ layouts) are provided through the project page at \url{https://shell-oss.github.io/cando/}. The test-set ground truth is intentionally withheld to support a hosted online evaluation server.

\item {\bf Experimental setting/details}
    \item[] Question: Does the paper specify all the training and test details (e.g., data splits, hyperparameters, how they were chosen, type of optimizer) necessary to understand the results?
    \item[] Answer: \answerYes{}
    \item[] Justification: The backbone models and GPU hardware are reported in Section~\ref{sec:experiments}; training and configuration details are provided in Appendix~\ref{app:experiments}. Evaluation metrics for our proposed benchmark are defined precisely in Section~\ref{sec:dataset}. Public-benchmark task definitions follow prior work.

\item {\bf Experiment statistical significance}
    \item[] Question: Does the paper report error bars suitably and correctly defined or other appropriate information about the statistical significance of the experiments?
    \item[] Answer: \answerYes{}
    \item[] Justification: Table~\ref{tab:cando_test_results} reports means and standard deviations over the 100-layout \textbf{ALPS-Bench} test set. Appendix~\ref{app:significance-testing} additionally reports paired two-sided Wilcoxon tests with Holm correction and paired-bootstrap $95\%$ confidence intervals for AGG, CS, and RE.

\item {\bf Experiments compute resources}
    \item[] Question: For each experiment, does the paper provide sufficient information on the computer resources (type of compute workers, memory, time of execution) needed to reproduce the experiments?
    \item[] Answer: \answerYes{}
    \item[] Justification: Hardware is summarized in Section~\ref{sec:experiments}; per-trial latency, parallel batch wall time, and local GPU-hours are reported in Appendix~\ref{app:compute}.
    
\item {\bf Code of ethics}
    \item[] Question: Does the research conducted in the paper conform, in every respect, with the NeurIPS Code of Ethics \url{https://neurips.cc/public/EthicsGuidelines}?
    \item[] Answer: \answerYes{}
    \item[] Justification: The research concerns automated layout design for operational facilities and raises no ethical concerns. The data does not include any personal identifiable information. The NeurIPS Code of Ethics has been reviewed.

\item {\bf Broader impacts}
    \item[] Question: Does the paper discuss both potential positive societal impacts and negative societal impacts of the work performed?
    \item[] Answer: \answerYes{}
    \item[] Justification: Potential positive and negative societal impacts are discussed in Appendix~\ref{app:broader-impact}.

\item {\bf Safeguards}
    \item[] Question: Does the paper describe safeguards that have been put in place for responsible release of data or models that have a high risk for misuse (e.g., pre-trained language models, image generators, or scraped datasets)?
    \item[] Answer: \answerNA{}
    \item[] Justification: The paper does not release a generative model with significant risk of misuse. \texttt{CANDO} is a task-specific layout design optimization framework and \textbf{ALPS-Bench} contains geometric facility layouts without personally identifiable or sensitive information.

\item {\bf Licenses for existing assets}
    \item[] Question: Are the creators or original owners of assets (e.g., code, data, models), used in the paper, properly credited and are the license and terms of use explicitly mentioned and properly respected?
    \item[] Answer: \answerYes{}
    \item[] Justification: All existing assets (PubLayNet, RICO, PKU-PosterLayout, Qwen3.6, Qwen3-VL-8B-Instruct, GPT-5.2, LayoutDM, and LayoutCoT) are properly cited, and their licenses and terms of use are listed in Appendix~\ref{app:licenses}.

\item {\bf New assets}
    \item[] Question: Are new assets introduced in the paper well documented and is the documentation provided alongside the assets?
    \item[] Answer: \answerYes{}
    \item[] Justification: \textbf{ALPS-Bench} ($1,000$ annotated layouts released alongside a design manual) is a new asset introduced in the paper. Its annotation format and structure are described in Appendices~\ref{app:annotation-format} and \ref{app:data-statistics}.

\item {\bf Crowdsourcing and research with human subjects}
    \item[] Question: For crowdsourcing experiments and research with human subjects, does the paper include the full text of instructions given to participants and screenshots, if applicable, as well as details about compensation (if any)? 
    \item[] Answer: \answerNA{}
    \item[] Justification: Annotation was performed by professional annotators working from original CAD drawings, not via a crowdsourcing platform. No human subjects research was conducted.

\item {\bf Institutional review board (IRB) approvals or equivalent for research with human subjects}
    \item[] Question: Does the paper describe potential risks incurred by study participants, whether such risks were disclosed to the subjects, and whether Institutional Review Board (IRB) approvals (or an equivalent approval/review based on the requirements of your country or institution) were obtained?
    \item[] Answer: \answerNA{}
    \item[] Justification: The paper does not involve research with human subjects and does not require IRB approval.

\item {\bf Declaration of LLM usage}
    \item[] Question: Does the paper describe the usage of LLMs if it is an important, original, or non-standard component of the core methods in this research? Note that if the LLM is used only for writing, editing, or formatting purposes and does \emph{not} impact the core methodology, scientific rigor, or originality of the research, declaration is not required.
    \item[] Answer: \answerYes{}
    \item[] Justification: LLMs are a central and original component of the \texttt{CANDO} framework. Qwen and GPT serve as backbone models (Section~\ref{sec:experiments}) for all of the proposed solution's agents. Their roles, versions, and usage are described throughout Section~\ref{sec:methodology}.

\end{enumerate}

\end{document}